\documentclass[english]{article}
\usepackage[T1]{fontenc}
\usepackage[utf8]{inputenc}
\usepackage[letterpaper]{geometry}
\usepackage{color}
\usepackage{babel}
\usepackage{verbatim}
\usepackage{float}
\usepackage{booktabs}
\usepackage{url}
\usepackage{amsmath}
\usepackage{amsthm}
\usepackage{amssymb}
\usepackage{graphicx}
\usepackage[unicode=true,pdfusetitle,
 bookmarks=true,bookmarksnumbered=false,bookmarksopen=false,
 breaklinks=false,pdfborder={0 0 1},backref=page,colorlinks=true]
 {hyperref}

\makeatletter

\providecommand{\tabularnewline}{\\}
\floatstyle{ruled}
\newfloat{algorithm}{tbp}{loa}
\providecommand{\algorithmname}{Algorithm}
\floatname{algorithm}{\protect\algorithmname}

\numberwithin{equation}{section}
\numberwithin{figure}{section}

\usepackage{tikz}
\usetikzlibrary{calc,matrix,fit}
\tikzset{pics/grid/.style={code={\tikzset{wonderich/.cd,#1}
    \def\pv##1{\pgfkeysvalueof{/tikz/wonderich/##1}}%
    \draw[thick] (0.1,0.1) grid (\pv{nx}+0.9,\pv{ny}+0.9);}},
pics/nodes/.style={code={\tikzset{wonderich/.cd,#1}
    \def\pv##1{\pgfkeysvalueof{/tikz/wonderich/##1}}%   
    \path  foreach \X in {1,...,\pv{nx}} {
      foreach \Y in {1,...,\pv{ny}} { 
      (\X,\Y)node[minimum size=\pv{r},style/.expanded=\pv{style},inner sep=0pt]{}} };
}},
wonderich/.cd,nx/.initial=4,ny/.initial=4,r/.initial=4pt,
    style/.initial={circle,fill}}
\usepackage[braket, qm]{qcircuit}

\usepackage{pgfplots}
\usepackage{tikzscale}

\@ifundefined{showcaptionsetup}{}{%
 \PassOptionsToPackage{caption=false}{subfig}}
\usepackage{subfig}
\makeatother

\usepackage{listings}

\begin{document}
\title{Tensor Networks for Simulating Quantum Circuits on FPGAs}
\author{Maksim Levental}
\maketitle
\begin{abstract}
Most research in quantum computing today is performed against simulations
of quantum computers rather than true quantum computers. Simulating
a quantum computer entails implementing all of the unitary operators
corresponding to the quantum gates as tensors. For high numbers of
qubits, performing tensor multiplications for these simulations becomes
quite expensive, since $N$-qubit gates correspond to $2^{N}$-dimensional
tensors. One way to accelerate such a simulation is to use field programmable
gate array (FPGA) hardware to efficiently compute the matrix multiplications.
Though FPGAs can efficiently perform tensor multiplications, they
are memory bound, having relatively small block random access memory.
One way to potentially reduce the memory footprint of a quantum computing
system is to represent it as a tensor network; tensor networks are
a formalism for representing compositions of tensors wherein economical
tensor contractions are readily identified. Thus we explore tensor
networks as a means to reducing the memory footprint of quantum computing
systems and broadly accelerating simulations of such systems.
\end{abstract}
\global\long\def\twovec#1#2{\begin{pmatrix}#1\\
 #2 
\end{pmatrix}}%

\global\long\def\C{\mathbb{C}}%

\global\long\def\R{\mathbb{R}}%

\global\long\def\bbra#1{\bra{#1}}%

\global\long\def\kket#1{\ket{#1}}%

\tableofcontents{}

\section{Introduction}

Quantum computing (QC) refers to the manipulation and exploitation
of properties of quantum mechanical (QM) systems to perform computation.
QM systems exhibit properties such as superposition and entanglement
and clever \textit{quantum algorithms} operate on these systems to
perform general computation. Unsurprisingly, the technique was intially
conceived of as a way to simulate physical systems themselves:
\begin{quotation}
``\ldots{} {[}N{]}ature isn't classical, dammit, and if you want to
make a simulation of nature, you'd better make it quantum mechanical,
and by golly it's a wonderful problem, because it doesn't look so
easy.''
\end{quotation}
This closing remark from the keynote at the 1\textsuperscript{st}
Physics of Computation Conference in 1981, delivered by the late Richard
Feynman \cite{feynman1982simulating}, succinctly, but accurately,
expresses that initial goal of quantum computing. Although modeling
and simulating physical systems on quantum computers remains a thriving
area of research we narrow our focus here to QC as it pertains to
solving general computational problems. Such problems include unstructured
search \cite{10.1145/237814.237866}, integer factorization \cite{365700},
combinatorial optimization \cite{farhi2014quantum}, and many others.
It is conjectured that some quantum algorithms enable quantum computers
to exceed the computational power of classical computers \cite{Zhong1460}. 

QC systems are composed of so-called quantum bits, or \textit{qubits},
that encode initial and intermediate states of computations. Transformations
between states are effected by time-reversible transforms, called
\textit{unitary} \textit{operators.} A formalism for representing
quantum computation is the \textit{quantum circuit} formalism, where
semenatically related collections of $N$ qubits are represented as
\textit{registers} and transformations are represented as \textit{gates},
connected to those registers by \textit{wires}, and applied in sequence.
As already mentioned, in hardware, quantum circuits correspond to
physical systems that readily exhibit quantum mechnical properties;
examples of physical qubits include transmons, ion traps and topological
quantum computers \cite{NAP25196}. Current state of the art QC systems
are termed Noisy Intermediate-Scale Quantum (NISQ) systems. Such systems
are characterized by moderate quantities of physical qubits (50-100)
but relatively few logical qubits (i.e. qubits robust to inteference
and noise). Due to these limitations (and, more broadly, the relative
scarcity of functioning QC systems), most research on quantum algorithms
is performed with the assistance of simulators of QC systems. Such
simulators perform simulations by representing $N$-qubit circuits
as $2^{N}$-dimensional complex vectors and transformations on those
vectors as $2^{N}$-dimensional complex matrix-vector multiplication.
Naturally, due to this exponential growth, naively executing such
simulations quickly become infeasible for $N>50$ qubits \cite{pednault2020paretoefficient},
both due to memory constraints and compute time. 

It's the case that matrices are a subclass of a more general mathematical
object called a \textit{tensor} and composition of matrices can be
expressed as \textit{tensor contraction}.\textit{ Tensor networks}
(TNs) are decompositions (i.e. factorizations) of very high-dimensional
tensors into networks (i.e. graphs) of low-dimensional tensors. TNs
have been successfullly employed in reducing the memory requirements
of simulations of QC systems \cite{pednault2020paretoefficient}.
The critical feature of tensor networks that make them useful for
QC is the potential to perform tensor contractions on the low-dimensional
tensors in an order such that, ultimately, the memory and compute
time requirements are lower than for the traditional representation.
Existing applications of TNs to quantum circuits focus primarily on
memory constraints on general purpose computers \cite{Fried_2018}
and distributed environments \cite{McCaskey_2018}. 

FPGAs are known to be performant for matrix multiplication uses cases
\cite{10.1145/3020078.3021740}. Though FPGAs typically run at lower
clock speeds (100-300MHz) than either conventional processors or even
graphics processors they, nonetheless, excel at latency constrained
computations owing to their fully ``synchronous'' nature (all modules
in the same \textit{clock domain} execute simultaneously). At first
glance FPGAs seem like a suitable platform for performant simulation
of quantum systems when runtime is of the essence. Unfortunately,
RAM is one of the more limited resources on an FPGA and therefore
it becomes necessary to explore memory reduction strategies for simulations
(as well as runtime reduction strategies). Hence, we explore tensor
networks as a means of reducing the memory footprint of quantum circuits
with particular attention to dimensions of the designs space as they
pertain to deployment to FPGAs. 

The remainder of this report is organized as follows: section \ref{sec:Background}
covers the necessary background wherein subsection \ref{sec:Quantum-Circuits}
very briefly reviews quantum computation and quantum circuits (with
particular focus on aspects that will be relevant for tensor networks
and FPGAs), section \ref{sec:Tensor-Networks} defines tensors and
tensor networks fairly rigorously and discusses algorithms for identifying
optimal contraction orders, section \ref{sec:FPGAs} discusses the
constraints imposed by virtue of deploying to FPGA, section \ref{sec:Implementation}
describes our implementation of TNs on FPGAs, section \ref{sec:Evaluation}
reports our results on various random circuits, and section \ref{sec:Conclusion}
concludes with future research directions.

\section{Background\label{sec:Background}}

\subsection{Quantum Computing\label{sec:Quantum-Circuits}}

We very (very) quickly review quantum computing and quantum circuits
as they pertain to our project. For a much more pedagogically sound
introduction consult \cite{j2020quantum}. As already alluded to,
quantum computing exploits properties of quantum mechanical systems
in order to perform arbitrary computation. The fundamental unit of
quantum computation is a qubit, defined as two-dimensional quantum
system with state vector $\psi$ an element of a Hilbert space\footnote{A Hilbert space $H$ is a vector space augmented with an inner product
such that, with respect to the metric induced by that inner product,
all Cauchy sequences converge.} $H$:
\[
\psi:=\alpha\twovec 10+\beta\twovec 01\equiv\twovec{\alpha}{\beta}
\]
where $\alpha,\beta\in\mathbb{\C}$ and $\left|\alpha\right|^{2}+\left|\beta\right|^{2}=1$.
This exhibits the superposition property of the qubit\footnote{We say that the qubit is in a superposition of the basis vectors/states.}
in that the squares of the coefficients are the probabilities of measuring
the system in the corresponding basis state. Collections of qubits
have state vectors that represented by the \textit{tensor product
}of the individual states of each qubit; for example, two qubits $\psi,\phi$
have state vector
\[
\psi\otimes\phi:=\twovec{\alpha}{\beta}\otimes\twovec{\alpha'}{\beta'}\equiv\begin{pmatrix}\alpha\alpha'\\
\alpha\beta'\\
\beta\alpha'\\
\beta\beta'
\end{pmatrix}
\]
where the second $\otimes$ is the Kronecker product and $\alpha\alpha'$
indicates conventional complex multiplication. Note that the basis
relative to which $\psi\otimes\phi$ is represented is the standard
basis for $\C^{4}$ and thus we observe exponential growth in the
size of the representation of an $N$-qubit system. An alternative
notation for state vectors is Dirac notation; for example, for a single
qubit
\[
\ket{\psi}\equiv\alpha\kket 0+\beta\kket 1
\]
and a 2-qubit system
\begin{align*}
\ket{\psi}\otimes\ket{\phi} & \equiv\left(\alpha\kket 0+\beta\kket 1\right)\otimes\left(\alpha'\kket 0+\beta'\kket 1\right)\\
 & \equiv\alpha\alpha'\kket 0\otimes\kket 0+\alpha\beta'\kket 0\otimes\kket 1+\beta\alpha'\kket 1\otimes\kket 0+\beta\beta'\kket 1\otimes\kket 1\\
 & \equiv\alpha\alpha'\kket 0\kket 0+\alpha\beta'\kket 0\kket 1+\beta\alpha'\kket 1\kket 0+\beta\beta'\kket 1\kket 1\\
 & \equiv\alpha\alpha'\kket{00}+\alpha\beta'\kket{01}+\beta\alpha'\kket{10}+\beta\beta'\kket{11}\\
 & \equiv\alpha\alpha'\kket 0+\alpha\beta'\kket 1+\beta\alpha'\kket 2+\beta\beta'\kket 3
\end{align*}
where in the last line we've used the decimal representation for the
bit strings identifying the basis states. Of particular import for
QC are the \textit{entangled }or \textit{bell states}; they correspond
to multi-qubit states, such as 
\[
\ket{\xi}=\frac{1}{\sqrt{2}}\ket{00}+\frac{1}{\sqrt{2}}\ket{11}
\]
that cannot be ``factored'' into component states\footnote{$\xi$ cannot be factored because there is no solution to the set
of equations (for $\alpha,\alpha',\beta,\beta'$)
\[
\alpha\alpha'=\frac{1}{\sqrt{2}},\quad\alpha\beta'=0,\quad\beta\alpha'=0,\quad\beta\beta'=\frac{1}{\sqrt{2}}
\]
}. Then, naturally, changes in qubit states are represented as unitary\footnote{A matrix $U$ is unitary iff $UU^{\dagger}=U^{\dagger}U=I$, i.e.
it is its own Hermitian conjugate or more simply if it is ``self-inverse''.} matrices $U$; for example
\[
\psi'=U\psi=\begin{pmatrix}U_{00} & U_{01}\\
U_{10} & U_{11}
\end{pmatrix}\twovec{\alpha}{\beta}=\twovec{U_{00}\alpha+U_{01}\beta}{U_{10}\alpha+U_{11}\beta}
\]
Matrix representations of transformations on multi-qubit states are
constructed using the Kronecker product on the individual matrix representations;
for example

\[
U\otimes V:=\begin{pmatrix}U_{00}V & U_{01}V\\ U_{10}V & U_{11}V \end{pmatrix}:=\begin{pmatrix}
 U_{00} V_{00} & U_{00} V_{01} & U_{01} V_{00} & U_{01} V_{01} \\
 U_{00} V_{10} & U_{00} V_{11} & U_{01} V_{10} & U_{01} V_{11} \\
 U_{10} V_{00} & U_{10} V_{01} & U_{11} V_{00} & U_{11} V_{01} \\
 U_{10} V_{10} & U_{10} V_{11} & U_{11} V_{10} & U_{11} V_{11}
\end{pmatrix}
\]Here we see again an exponential growth in representation size as
a function of number of qubits. 

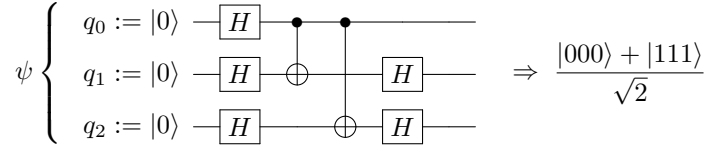
\begin{figure}
\begin{equation*}
    \psi\begin{cases}\qquad\qquad\quad
	\Qcircuit @C=1.0em @R=0.7em {
	 	\lstick{ q_0 := \ket{0} } & \gate{H} & \ctrl{1} & \ctrl{2} & \qw & \qw & \qw
		 \\
	 	\lstick{ q_1 := \ket{0}  } & \gate{H} & \targ & \qw & \gate{H} & \qw & \qw 
		 \\
	 	\lstick{ q_2 := \ket{0}  } & \gate{H} & \qw & \targ & \gate{H} & \qw & \qw
		 \\
	 	%\lstick{c:} & {/_{_{1}}} \cw & \cw & \cw & \cw & \cw & \cw\\
	 }
	\end{cases} 
	\Rightarrow\; \frac{\ket{000} + \ket{111}}{\sqrt{2}}
\end{equation*}
\centering{}\caption{Quantum Circuit representing 3-qubit $q_{0},q_{1},q_{2}$ entanglement
effected by application of successive Hadamard gates.\label{fig:Quantum-Circuit-representing-1}}
\end{figure}

As already alluded to, quantum circuits are a formalism for representing
quantum computation in general and algorithms designed for quantum
computers in particular. In the quantum circuit formalism qubit states
are represented by wires and unitary transformations are represented
by gates (see figure \ref{fig:Quantum-Circuit-representing-1}), much
like classical combinational logic circuits might be, though, whereas
combinational logic is ``memoryless''\footnote{The output of a combinational logic circuit at any time is only a
function of the elements of the circuit and its inputs.}, sequences of quantum gates specified by a quantum circuit do in
fact connote the evolution (in time) of the qubits. In addition quantum
gates, as opposed to classical gates, are necessarily reversible and
hence there are no quantum analogs to some classical gates such as
NOT and OR.

\subsection{Tensors and Tensor Networks\label{sec:Tensor-Networks}}

We quickly define tensors and tensor networks and then move on to
tensor network methods for simulating quantum circuits.

\subsubsection{Tensors}

One definition of a tensor\footnote{There are several more at varying levels of mathematical sophistication.
Chapter 14 of \cite{roman2007advanced} is the standard reference.
Ironically, it is this author's opinion that one should shy away from
physics oriented expositions on tensors.} $T$ is as an element of a tensor product space\footnote{The collection of tensor products of elements of the component spaces
quotiented by an equivalence relation.}:
\[
T\in\underbrace{V\otimes\cdots\otimes V}_{p{\text{ copies}}}\otimes\underbrace{V^{*}\otimes\cdots\otimes V^{*}}_{q{\text{ copies}}}
\]
where $V^{*}$ is dual\footnote{The dual space to a vector space $V$ is the vector $V^{*}$ consisting
of linear maps $f:V\rightarrow\R$. The dual basis of the dual space
consists of $f_{i}$ such that $f_{i}\left(\mathbf{e}_{i}\right)=\delta_{ij}$.
It is convention to write $f_{i}$ as $\mathbf{e}^{i}$ (note the
superscript index).} to $V$. Then $T$, in effect, acts a multilinear map
\[
{\displaystyle T:\underbrace{V^{*}\times\dots\times V^{*}}_{p{\text{ copies}}}\times\underbrace{V\times\dots\times V}_{q{\text{ copies}}}\rightarrow\R}
\]
by ``applying'' $p$ elements from $V$ to $p$ elements of $V^{*}$
and $q$ elements from $V^{*}$ to $q$ elements of $V$. Note the
swapping of the orders of $V,V^{*}$ in both the definitions and the
description. $T$'s coordinate basis representation

\begin{equation}
{\displaystyle T\equiv T_{j_{1}\dots j_{q}}^{i_{1}\dots i_{p}}\;\mathbf{e}_{i_{1}}\otimes\cdots\otimes\mathbf{e}_{i_{p}}\otimes\mathbf{e}^{j_{1}}\otimes\cdots\otimes\mathbf{e}^{j_{q}}}\label{eq:tensor_coord}
\end{equation}
is determined by its evaluation on each set of bases

\[
{\displaystyle T_{j_{1}\dots j_{q}}^{i_{1}\dots i_{p}}:=T\left(\mathbf{e}^{i_{1}},\ldots,\mathbf{e}^{i_{p}},\mathbf{e}_{j_{1}},\ldots,\mathbf{e}_{j_{q}}\right)}
\]
The pair $\left(p,q\right)$ is called the \textit{type }or\textit{
valence} of \textbf{$T$} while $\left(p+q\right)$ is the \textit{order}
of the tensor. \textbf{Note that we do not use rank to mean either
of these things}\footnote{The \textit{rank} of a tensor is the minimum number of distinct basis
tensors necessary to define it; the tensor in eqn. (\ref{eq:tensor_coord})
is in fact a rank 1 tensor. The definition is a generalization of
the rank of a matrix (which, recalling, is the dimension of its column
space, i.e. number of basis elements). Despite this obvious, reasonable
definition for rank, one should be aware that almost all literature
in this area of research uses rank to mean order.}. Furthermore, eqn. (\ref{eq:tensor_coord}) in fact represents a
linear sum of basis elements, as it employs Einstein summation convention\footnote{Repeated indices in juxtapose position indicate summation $a_{i}b^{i}:=\sum_{i}a[i]b[i]$.}.
Note we make liberal use of summation convention in the following
but occasionally use explicit sums when it improves presentation (i.e.
when we would like to emphasize a particular contraction).

There are two important operations on tensors we need to define. Firstly,
we can form the tensor product $Z$ of two tensors $T,W$, of types
$\left(p,q\right),\left(r,s\right)$ respectively, to obtain a tensor
of type $\left(p+r,q+s\right)$:
\begin{align*}
Z & :=T\otimes W\\
 & \;=\left(T_{j_{1}\dots j_{q}}^{i_{1}\dots i_{p}}\;\mathbf{e}_{i_{1}}\otimes\cdots\otimes\mathbf{e}_{i_{p}}\otimes{\mathbf{e}}^{j_{1}}\otimes\cdots\otimes{\mathbf{e}}^{j_{q}}\right)\otimes\left(W_{l_{1}\dots l_{s}}^{k_{1}\dots k_{r}}\;\mathbf{e}_{k_{1}}\otimes\cdots\otimes\mathbf{e}_{k_{r}}\otimes{\mathbf{e}}^{l_{1}}\otimes\cdots\otimes{\mathbf{e}}^{l_{s}}\right)\\
 & \;=\left(T_{j_{1}\dots j_{q}}^{i_{1}\dots i_{p}}W_{l_{1}\dots l_{s}}^{k_{1}\dots k_{r}}\;\mathbf{e}_{i_{1}}\otimes\cdots\otimes\mathbf{e}_{i_{p}}\otimes\mathbf{e}_{k_{1}}\otimes\cdots\otimes\mathbf{e}_{k_{r}}\otimes{\mathbf{e}}^{j_{1}}\otimes\cdots\otimes{\mathbf{e}}^{j_{q}}\otimes{\mathbf{e}}^{l_{1}}\otimes\cdots\otimes{\mathbf{e}}^{l_{s}}\right)\\
 & :=Z_{j_{1}\dots j_{q+s}}^{i_{1}\dots i_{p+r}}\;\mathbf{e}_{i_{1}}\otimes\cdots\otimes\mathbf{e}_{i_{p+r}}\otimes{\mathbf{e}}^{j_{1}}\otimes\cdots\otimes{\mathbf{e}}^{j_{q+s}}
\end{align*}
\begin{comment}
The attentive reader should notice that the coordinate representation
of two tensors is 
\end{comment}
Despite it being obvious, its important to note that the tensor product
$Z$ produces a tensor of order $\left(p+r+q+s\right)$, i.e. higher
than either of the operands. On the contrary, \textit{tensor contraction}
reduces the order of a tensor. We define the contraction $Y$ of type
$\left(a,b\right)$ of a tensor $T$ to be the ``pairing'' of the
$a$th and $b$th bases: 
\begin{align*}
Y & :=T_{j_{1}\dots j_{q}}^{i_{1}\dots i_{p}}\;\mathbf{e}_{i_{1}}\otimes\cdots\otimes\mathbf{e}_{i_{a-1}}\otimes\mathbf{e}_{i_{a+1}}\otimes\cdots\otimes\mathbf{e}_{i_{p}}\otimes\left(\mathbf{e}_{i_{a}}\cdot\mathbf{e}^{j_{b}}\right)\otimes\mathbf{e}^{j_{1}}\otimes\cdots\otimes\mathbf{e}^{j_{b-1}}\otimes\mathbf{e}^{j_{b+1}}\otimes\cdots\otimes\mathbf{e}^{j_{q}}\\
 & \;=T_{j_{1}\dots j_{q}}^{i_{1}\dots i_{p}}\delta_{i_{a}}^{j_{b}}\;\mathbf{e}_{i_{1}}\otimes\cdots\otimes\mathbf{e}_{i_{a-1}}\otimes\mathbf{e}_{i_{a+1}}\otimes\cdots\otimes\mathbf{e}_{i_{p}}\otimes\mathbf{e}^{j_{1}}\otimes\cdots\otimes\mathbf{e}^{j_{b-1}}\otimes\mathbf{e}^{j_{b+1}}\otimes\cdots\otimes\mathbf{e}^{j_{q}}\qquad\left(\text{since }\mathbf{e}_{i}\cdot\mathbf{e}^{j}=\delta_{i}^{j}\right)\\
 & \;=\sum_{j_{b}}T_{j_{1}\dots j_{b}\dots j_{q}}^{i_{1}\dots j_{b}\dots i_{p}}\;\mathbf{e}_{i_{1}}\otimes\cdots\otimes\mathbf{e}_{i_{a-1}}\otimes\mathbf{e}_{i_{a+1}}\otimes\cdots\otimes\mathbf{e}_{i_{p}}\otimes\mathbf{e}^{j_{1}}\otimes\cdots\otimes\mathbf{e}^{j_{b-1}}\otimes\mathbf{e}^{j_{b+1}}\otimes\cdots\otimes\mathbf{e}^{j_{q}}\\
 & :=Y_{j_{1}\dots i_{b-1}i_{b+1}\dots j_{q}}^{i_{1}\dots i_{a-1}i_{a+1}\dots i_{p}}\;\mathbf{e}_{i_{1}}\otimes\cdots\otimes\mathbf{e}_{i_{a-1}}\otimes\mathbf{e}_{i_{a+1}}\otimes\cdots\otimes\mathbf{e}_{i_{p}}\otimes\mathbf{e}^{j_{1}}\otimes\cdots\otimes\mathbf{e}^{j_{b-1}}\otimes\mathbf{e}^{j_{b+1}}\otimes\cdots\otimes\mathbf{e}^{j_{q}}
\end{align*}
where $\left(\cdot\right)$ means inner product. Notice that the order
of $Y$ is $\left(p-1,q-1\right)$. Finally notice that we can omit
writing out bases and just manipulate coordinates. We shall do as
such when it simplifies presentation.

As mentioned in the introduction, matrices can be represented as tensors;
for example, the two dimensional $N\times N$ matrix $M$ is taken
to be a tensor of type $\left(1,1\right)$ with basis representation
\[
{\displaystyle M\equiv M_{j}^{i}\;\mathbf{e}_{i}\otimes\mathbf{e}^{j}}
\]
where upper indices correspond to the row index and lower indices
correspond to the column index of the conventional matrix representation
and both range from $1$ to $N$. The attentive reader will notice
that the coordinate representation of the tensor product for type
$\left(1,1\right)$ tensors is exactly the Kronecker product for matrices.
Similarly, tensor contraction for type $\left(1,1\right)$ tensors
is the familiar matrix trace:

\[
M_{j}^{i}\left(\mathbf{e}_{i}\cdot\mathbf{e}^{j}\right)=M_{j}^{i}\delta_{i}^{j}=\sum_{i=1}^{N}M_{i}^{i}
\]
More usefully, we can express matrix-vector multiplication in terms
of tensor contraction; let\textit{ 
\[
\mathbf{x}:=\twovec{x^{1}}{x^{2}}\equiv x^{1}\mathbf{e}_{1}+x^{2}\mathbf{e}_{2}\equiv x^{i}\mathbf{e}_{i}
\]
}where we switch to valence index notation in the column vector for
closer affinity with tensor notation. Then it must be the case that
\[
\mathbf{y}=M\mathbf{x}=\left(M_{j}^{i}\mathbf{e}_{i}\otimes\mathbf{e}^{j}\right)\left(x^{k}\mathbf{e}_{k}\right)=\left(M_{j}^{i}x^{k}\mathbf{e}_{i}\right)\left(\mathbf{e}^{j}\cdot\mathbf{e}_{k}\right)=M_{j}^{i}x^{k}\delta_{k}^{j}\mathbf{e}_{i}=M_{j}^{i}x^{j}\mathbf{e}_{i}
\]
Letting $y^{i}:=M_{j}^{i}x^{j}$ we recognize conventional matrix-vector
multiplication. Employing tensor contraction in this way extends to
matrix-matrix multiplication (and tensor composition more broadly);
for two type $\left(1,1\right)$ tensors $M,L$ we can form the type
$\left(1,1\right)$ tensor $Z$ corresponding to matrix product $M\cdot L$
of $N\times N$ by first taking the tensor product

\[
Z_{lj}^{ik}:=M_{l}^{i}L_{j}^{k}
\]
The attentive reader will notice that the coordinate representation
of two tensors is exactly the Kronecker product of two matrices. Then
contracting along the off diagonal
\begin{equation}
Z_{j}^{i}:=Z_{kj}^{ik}=M_{k}^{i}L_{j}^{k}\equiv\sum_{k=1}^{N}M_{k}^{i}L_{j}^{k}\label{eq:matrix_mult}
\end{equation}
One can confirm that this is indeed conventional matrix multipliation
of two $N\times N$ matrices. In general, stated simply, when contracting
indices of a tensor product, contraction can be understood to be a
sum over shared indices.

\subsubsection{Tensor Networks}

\begin{comment}
\begin{figure}
\centering{}    \begin{tikzpicture}
     \draw[dashed]  pic{grid}
     (0.3,4.7) node{$\mathbf{S}\boldsymbol{'}$};
     \clip (0.35,0.35) rectangle (4.65,4.65);
     \path[rotate around={45:(45:2.5)},scale={sqrt(2)},
        shift={(0,-0.75)},transform shape]  pic{grid={nx=3,ny=3}}
        pic[red]{nodes={r=4pt,nx=3,ny=3}};
    \end{tikzpicture}\caption{Projected Entangled Pair State (PEPS) for a 3 \texttimes{} 3 lattice
with open boundary conditions.\label{fig:Projected-Entangled-Pair}}
\end{figure}
\end{comment}

\begin{figure}
\begin{centering}
\subfloat[Contraction \label{fig:Tensor-networks-demonstrating}]{\begin{centering}
\includegraphics[width=0.75\textwidth]{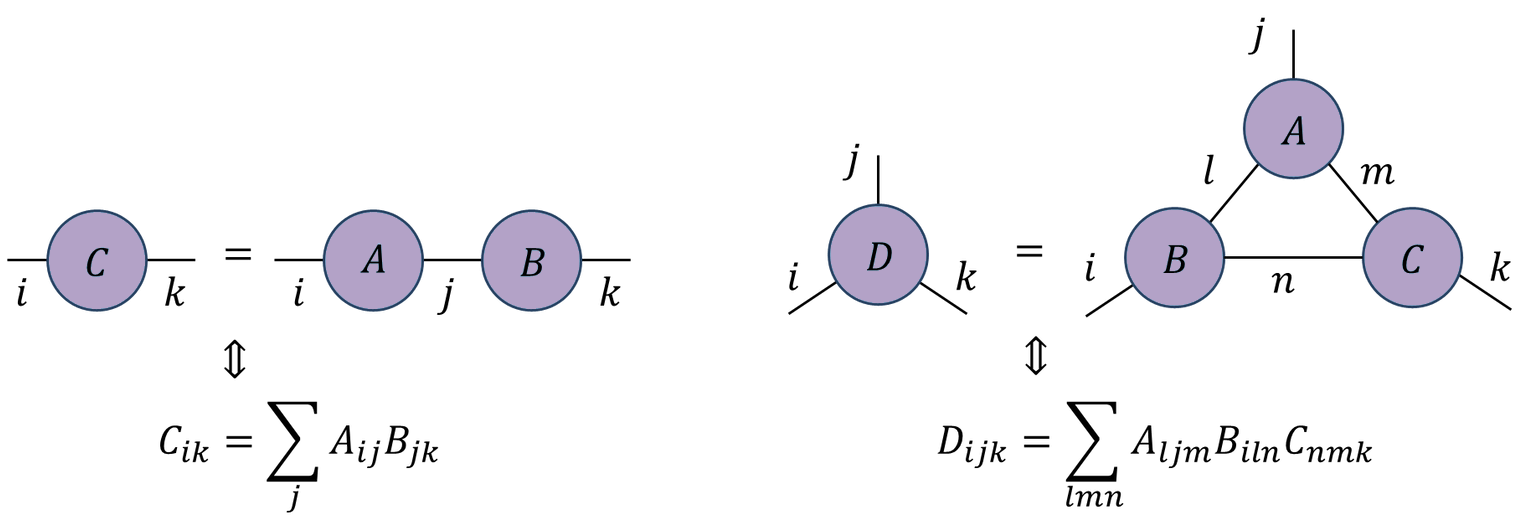}
\par\end{centering}
}
\par\end{centering}
\centering{}\smallskip{}
\subfloat[State vector representation\label{fig:State-vector-representation}]{\begin{centering}
\includegraphics[width=0.75\textwidth]{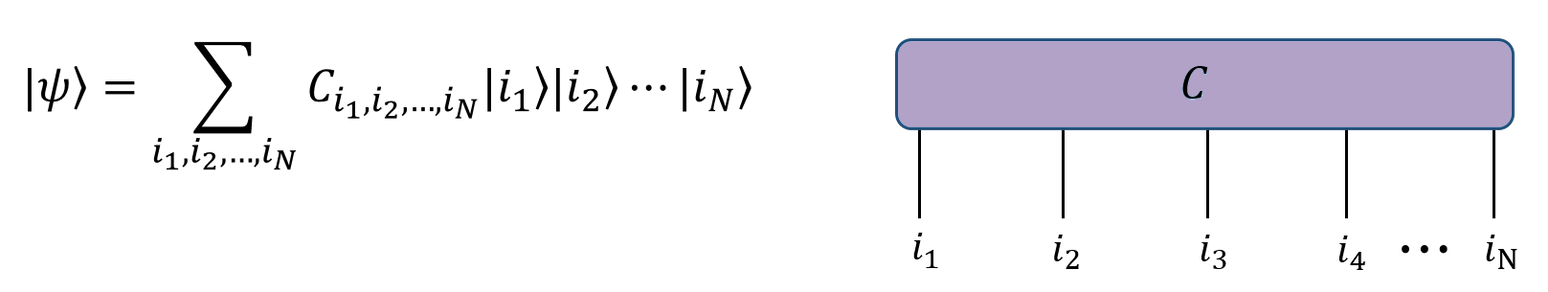}
\par\end{centering}
}\smallskip{}
\subfloat[State vector factorization\label{fig:State-vector-factorization}]{\begin{centering}
\includegraphics[width=0.75\textwidth]{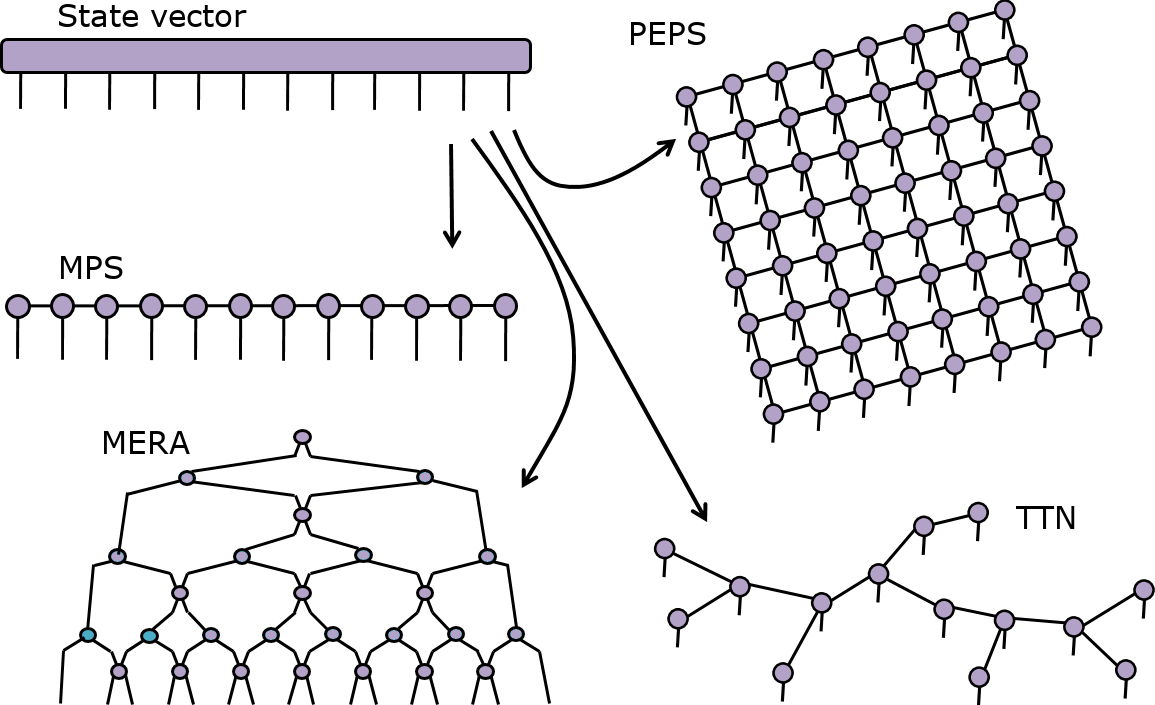}
\par\end{centering}
}\caption{Tensor network diagrammatic contraction. \cite{evenbly2019tensortrace}}
\end{figure}

Tensor networks (TNs) are a way to factor tensors with large orders
into networks of tensors with lower orders; since the number of parameters
a tensor consists of is exponential in the order of the tensor, smaller
order tensors are much preferrable computationally. They were first
used to study ground states of one dimensional quantum many-body systems
\cite{PhysRevLett.69.2863} but have since been applied in other areas
(such as machine learning \cite{glasser2019expressive}). TNs lend
themselves to a diagrammatic representation which can be used to reason
about such factorizations (figure \ref{fig:Tensor-networks-demonstrating}).
We will primarily be interested in TNs as a means to factoring the
state-vector of an $N$-qubit system (see figure \ref{fig:State-vector-representation})
\begin{equation}
\ket{\psi}:=\sum_{i_{1}i_{2}\dots i_{N}}C^{i_{1}i_{2}\dots i_{N}}\ket{i_{1}}\ket{i_{2}}\cdots\ket{i_{N}}\label{eq:state_vector}
\end{equation}
for which its common to propose an ansatz factorizations: 
\begin{itemize}
\item \textbf{Matrix Product States (MPS)} \cite{Kl_mper_1993}, which yields
factorization
\[
C^{i_{1}i_{2}\dots i_{N}}\equiv A_{j_{1}}^{i_{1}}A_{j_{2}}^{i_{1}j_{1}}\cdots A_{j_{N-1}}^{i_{N-1}j_{N-2}}A^{i_{N}j_{N-1}}
\]
where $j$ are called \textit{bond indices. }If each index $i$ has
dimension $d$ (i.e. takes on values 1 to $d$) then $C$ is specified
by $d^{N}$ parameters and can always be represented by an MPS factorization
$Ndm^{2}$ parameters, where $m:=d^{N/2}$ is the bond dimension.
While for this naive representation $d^{N}<Ndm^{2}$, in practice
$m$ is fixed to some moderate size such that $d^{N}>Ndm^{2}$ and
the MPS factorization functions as an approximation.
\item \textbf{Projected Entangled Pair States (PEPS)} \cite{Verstraete:2004cf},
which is a generalization of MPS to higher spatial dimensions, i.e.
TNs that correspond to lattices of contractions of tensors, which
themselves represent pairwise entangled quantum systems. Naturally,
such a series of contractions doesn't lend itself to being expressed
in traditional notation and thus we observe the power of tensor network
diagrams (see PEPS in figure \ref{fig:State-vector-factorization}).
\item \textbf{Tree Tensor Networks (TTN) }\cite{PhysRevA.74.022320}, a
further generalization where tensors are entangled (and therefore
contracted) hierarchically.\textbf{ }In fact TTNs bear the closest
resemblance to quantum circuits.
\item \textbf{Multi-scale Entanglement Renormalization Ansatz (MERA) }\cite{PhysRevLett.99.220405},
a specific type of TTN where the tensors are alternatingly unitaries
and isometries\footnote{A tensor, seen as a multlinear map, that preserves distances under
the ambient distance metric.}.
\end{itemize}

\subsubsection{TNs for Simulating Quantum Circuits}

Factoring eqn. (\ref{eq:state_vector}) is only the first step to
successfully simulating a quantum circuit. By representing some final
state as a tensor as well, and contracting across all indices (called
\textit{contracting the network}), we can calculate the amplitude
for that particular state. Since tensor contraction is associative\footnote{This can be observed by noting that summing is an associative operation
(or by analog with matrix-matrix multiplication).}, the order in which tensors are actually contracted is a ``hyperparameter''
of TN methods; finding the optimal contraction order, with respect
to accuracy (assuming some approximation has been made in constructing
the factorization), compute time, and memory requirements is critical. 

In particular we focus on contraction orders for TTNs as they most
closely resemble quantum circuits. For a TTN consisting of $N$ tensors
(corresponding to $N$ gates) with maximum order $p$, worst case,
we can see that contraction time takes $O\left(N\exp\left(O\left(p\right)\right)\right)$
since, in general, contracting across all indices of a pair of tensors
is exponential in their orders\footnote{Consider contracting two $\left(1,1\right)$ tensors (as in eqn. (\ref{eq:matrix_mult})),
i.e. two order 2 tensors, which effectively is matrix multplication
followed by trace. The complexity of this contraction is then $O\left(N^{2+1}+N\right)\equiv O\left(\exp\left(2\log N\right)\left(1+N\right)\right)$
(where $N$ here is the characteristic dimension of the matrix). Assuming
the ranges of all tensor indices is the same (i.e. $N$ is constant
across all tensors), for example $N=2$ as in the case of matrices
derived of unitary transformations operating on single qubits, we
recover the stated complexity. }. Markov et al. \cite{10.1137/050644756} showed that there in fact
exists a contraction ordering which results in a contraction time
of $O\left(N^{O\left(1\right)}\exp\left(O\left(\operatorname{tw}\left(G^{L}\right)\right)\right)\right)$
where $G^{L}$ is the line graph\footnote{A \textit{line graph }captures edge adjacency; given a graph $G$,
$G^{L}$ is defined such that each edge of $G$ corresponds to a vertex
of $G^{L}$ and two vertices are are connected in $G^{L}$ if the
edges in $G$ that they correspond to are adjacent on the same vertex
(in $G$). } of the tensor network and $\operatorname{tw}\left(G^{L}\right)$
is the tree-width\footnote{A\textit{ tree decomposition} of a graph $G$ is a tree $T$ and a
mapping from the vertices of $G$ into ``bags'' that satisfy the
following properties
\begin{enumerate}
\item Each vertex must appear in at least one bag.
\item For each edge in $G$, at least one bag must contain both of the vertices
it is adjacent on.
\item All bags containing a given vertex in $G$ must be connected in $T$.
\end{enumerate}
The width $w$ of a tree decomposition is the cardinality of the largest
bag (minus one). Finally the \textit{tree-width }of $G$ is the minimum
width over all possible tree decompositions. Intuitively, a graph
has low tree-width if it can be constructed by joining small graphs
together into a tree.} of $G^{L}$. %
\begin{comment}
Thus, in general, we can simulate arbitrary quantum circuits by
\begin{enumerate}
\item Constructing the TTN (with graph $G$) that corresponds to the quantum
circuit.
\item Computing a tree-decomposition of $G^{L}$ with width $\operatorname{tw}\left(G^{L}\right)$.
\item Finding a contraction ordering\footnote{Markov et al. also show that one can recover a contraction order from
the tree-decomposition of $G^{L}$ in polynomial time.} for the TTN and fully contracting to a single, real, number (i.e.
order 0 tensor).
\end{enumerate}
Note that, in general, computing the tree-width of a graph is NP-hard
\cite{Arnborg87complexityof} and Markov et al.'s results rely on
generating \textit{any} tree-decomposition that has a tree-width that
suits their needs (i.e. overall runtime complexity).
\end{comment}
{} For quantum circuits consisting of many few qubit gates this technique
produces a much more (runtime) efficient evaluation of the circuit;
indeed Markov et al. further show that any TTN corresponding to a
quantum circuit with $N$ gates, where the number of gates that act
on any pair of qubits is bounded by $r$, has contraction time $O\left(N^{O\left(1\right)}\exp\left(O\left(r\right)\right)\right)$.

Markov et al.'s results are not tight; their construction finds some
tree-decomposition with the correct corresponding tensor contraction
order that suits their aim (overall runtime complexity of the translation
from quantum circuit to TTN and the ultimate contraction). In reality
there are often contraction orders that are much more space and runtime
efficient. Though, in general problem is NP-hard \cite{Arnborg87complexityof},
for particular TTNs (corresponding to circuits) there are heuristics,
such as non-adjacent contractions \cite{pednault2020paretoefficient},
that produce more efficient orders. Alternatively, randomized search
and Bayesian optimization can be used to identify efficient contraction
orders \cite{Gray_2021,Fried_2018}.

\subsection{FPGAs\label{sec:FPGAs}}

\begin{figure}
\centering{}\includegraphics[width=0.5\textwidth]{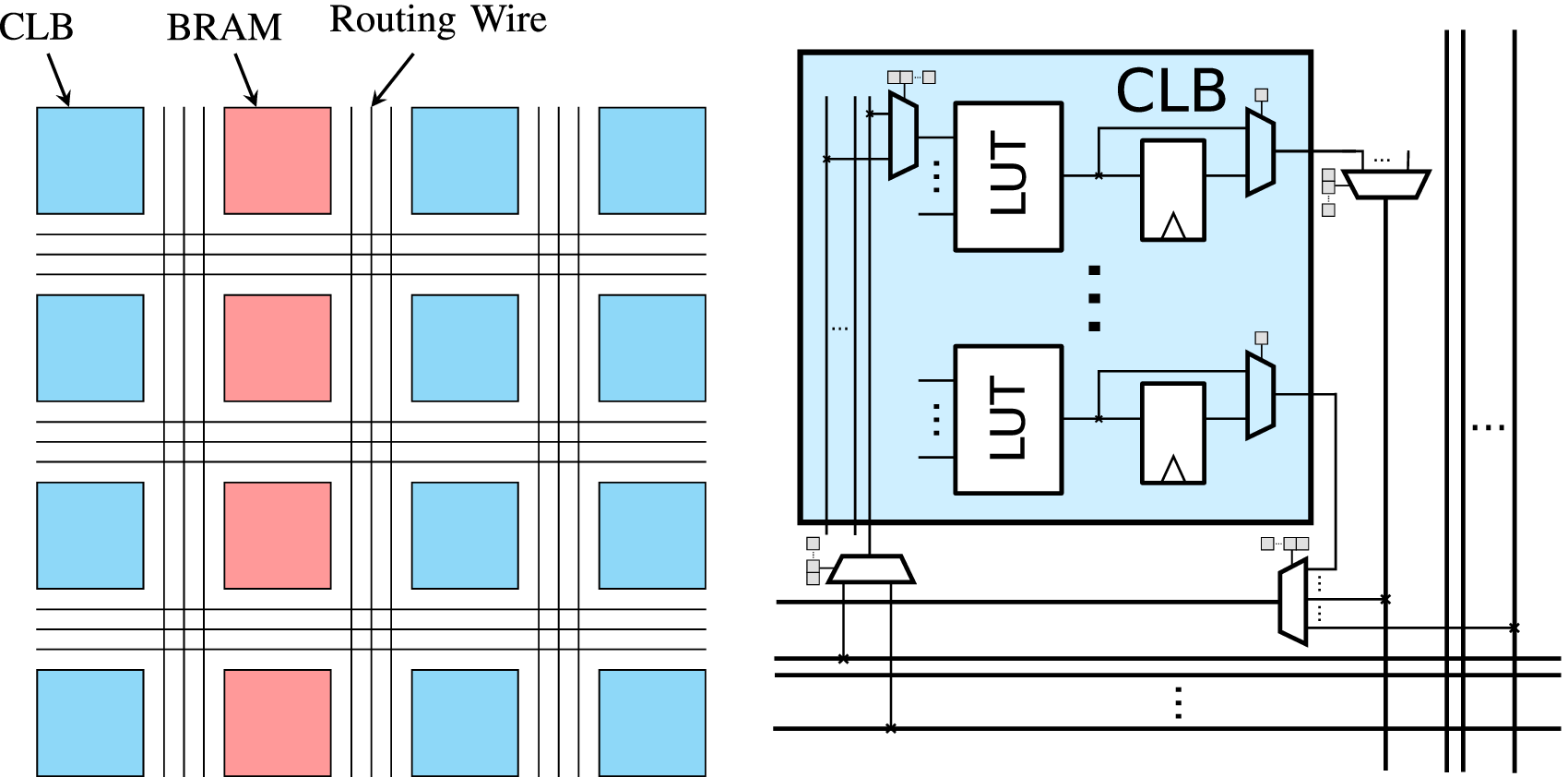}\caption{FPGA floorplan diagram \cite{9103284}.\label{fig:FPGA-floorplan-diagram}}
\end{figure}

A field-programmable gate array (FPGA) is a device designed to be
configured by a user into various arrangements of (classical) gates
and memory. FPGAs consist of arrays (hence the name) of configurable
logic blocks (CLBs), static ram (SRAM), and programmable busses that
connect CLBs and SRAM into various topologies (see figure \ref{fig:FPGA-floorplan-diagram}).
The CLBs typically contain arithmetic units (such as adders, multipliers,
and accumulators) and lookup tables (LUTs), that can be programmed
to represent truth tables for many boolean functions. Using hardware
description languages (such as VHDL or Verilog) designers specify
modules and compose them into circuits (also known as a \textit{dataflows})
that perform arbitrary computation. These circuits then go through
a \textit{place and route} procedure before ultimately being instantiated
on the FPGA as \textit{processing elements} (PEs) and connections
between PEs.

\begin{figure}
\centering{}\subfloat[Gemmini systolic array architecture. The processing elements (PEs)
are either of type Weight Stationary (WS) or Output Stationary (OS).
\cite{genc2019gemmini}.\label{fig:systolic-array-2}]{\centering{}\includegraphics[width=0.8\textwidth]{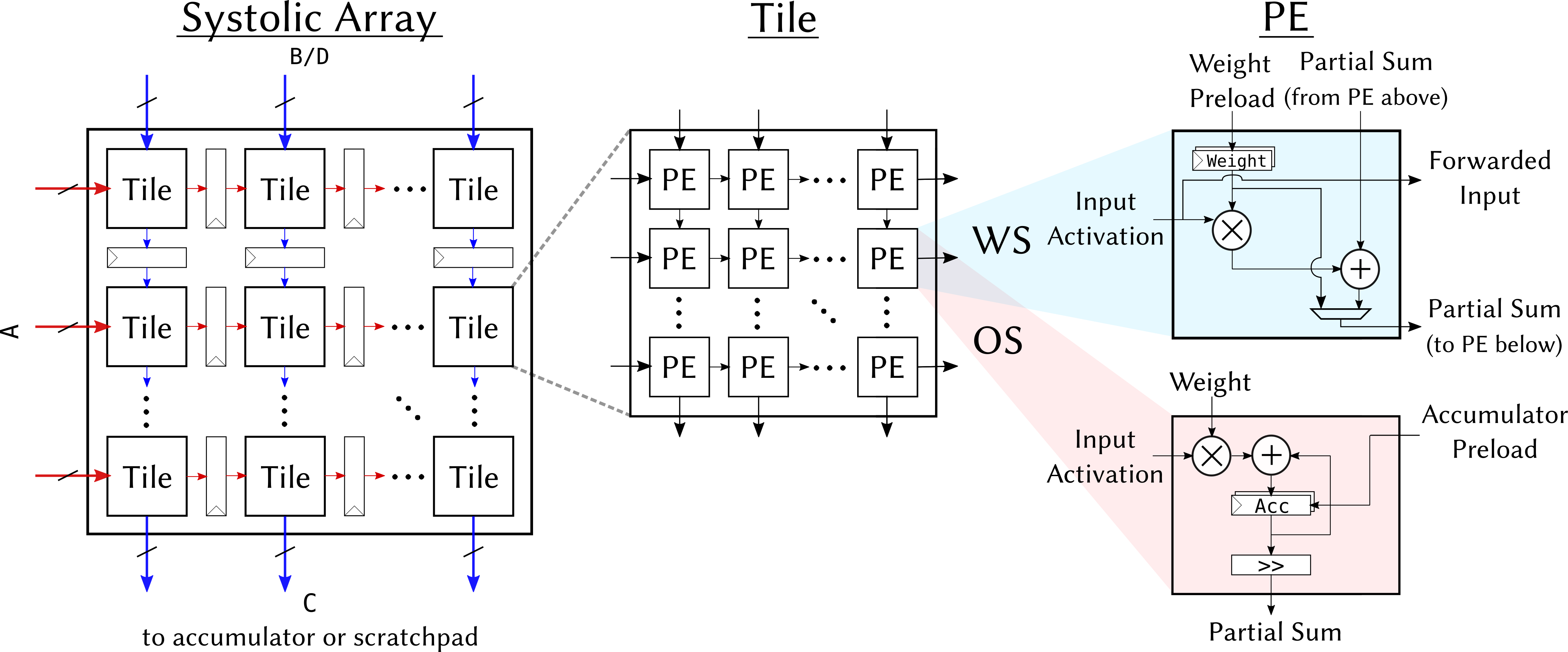}}\smallskip{}
\subfloat[Systolic array architecture implementing matrix multiplication. Input
matrices $A$ and $B$ stream by to produce output matrix $C$ via
successive mutiply-accumulate (MAC) operations. Note that $C$ remains
in the processing elements (i.e. this is a diagram of an OS architecture).
\cite{10.1007/978-3-030-05677-3_16}.\label{fig:systolic-array}]{\centering{}\includegraphics[width=0.8\textwidth]{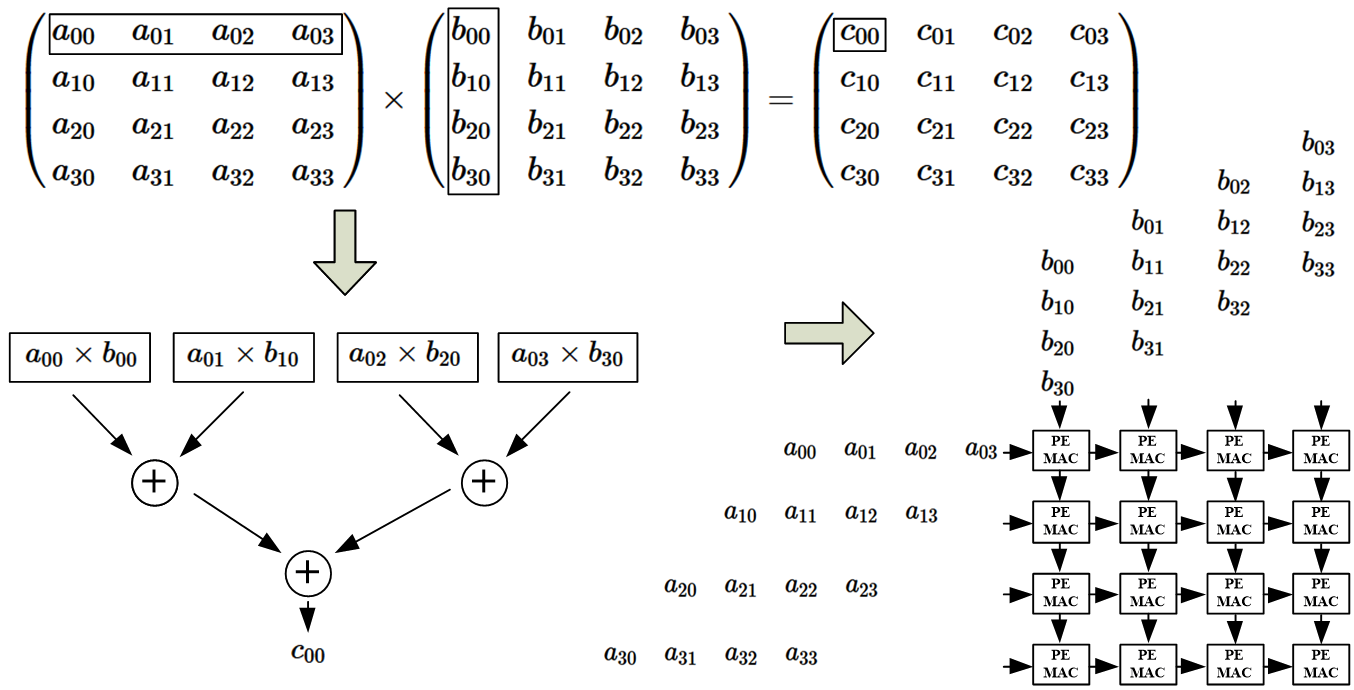}}\caption{Systolic arrays}
\end{figure}

While modules consisting purely of combinational logic compute their
outputs at the stated clock speed of the FPGA, inevitably I/O (i.e.
fetching data from memory) interleaved with such modules (otherwise
arranged into a pipeline architecture) creates pipeline stalls. Thus,
it's essential that FPGA designs are as compute bound as possible
(rather than I/O bound). In particular, we explore I/O minimal generalized
matrix multiplication (GEMM) \cite{10.1145/3373087.3375296} and other
\textit{systolic array} architectures \cite{10.1145/3431920.3439292,genc2019gemmini}.
A systolic architecture \cite{1653825} is a gridded, pipelined, array
of PEs that processes data as the data flows\footnote{The relationship to cardiovascular ``systolic'' is in association
with the flow of data into the array, akin to how blood flows through
the veins into the human heart.} through the array. Crucially, a systolic architecture propagates
partial results as well as input data through the pipeline (see figure
\ref{fig:systolic-array-2}). Systolic arrays are particularly suited
for I/O efficient matrix multiplication owing to the pipelining of
inputs (see figure \ref{fig:systolic-array}).

One remaining hurdle to simulating quantum compuations (i.e. carrying
out tensor contractions) on FPGAs is SRAM. The standard remediation
is to perform arithmetic with reduced precision\footnote{Germaine to this issue is the fact that arithmetic on FPGAs is typically
performed in fixed precision (via an integer representation), owing
to higher compute cost incurred for floating point arithmetic.}. There is evidence that suggests that simulations of quantum circuits,
of varying depths \cite{betelu2020limits}, are robust to reduced
precision computation as long as that loss of precision is uncorrelated
\cite{10.1145/3295500.3356155} i.e. insofar as it can be treated
as uncorrelated noise.

\section{Implementation\label{sec:Implementation}}

\begin{figure}
\begin{equation*}
	n\underbrace{\begin{cases}\qquad
    \Qcircuit @C=0.5em @R=1.0em @!R { 	 	\lstick{ {q}_{0} :  } & \gate{H} & \ctrl{1} & \gate{R_z} & \gate{R_x} & \qw & \ctrl{1} & \gate{R_z} & \gate{R_x} & \qw & \ctrl{1} & \gate{R_z} & \gate{R_x} & \qw & \ctrl{1} & \gate{R_z} & \gate{R_x} & \gate{H} & \qw & \qw & \qw && \cdots\\ 	 	\lstick{ {q}_{1} :  } & \gate{H} & \targ & \gate{R_z} & \ctrl{1} & \gate{R_x} & \targ & \gate{R_z} & \ctrl{1} & \gate{R_x} & \targ & \gate{R_z} & \ctrl{1} & \gate{R_x} & \targ & \gate{R_z} & \ctrl{1} & \gate{R_x} & \gate{H} & \qw & \qw && \cdots\\ 	 	\lstick{ {q}_{2} :  } & \gate{H} & \ctrl{1} & \gate{R_z} & \control\qw & \gate{R_x} & \ctrl{1} & \gate{R_z} & \control\qw & \gate{R_x} & \ctrl{1} & \gate{R_z} & \control\qw & \gate{R_x} & \ctrl{1} & \gate{R_z} & \control\qw & \gate{R_x} & \gate{H} & \qw & \qw && \cdots\\ 	 	\lstick{ {q}_{3} :  } & \gate{H} & \targ & \gate{R_z} & \ctrl{1} & \gate{R_x} & \targ & \gate{R_z} & \ctrl{1} & \gate{R_x} & \targ & \gate{R_z} & \ctrl{1} & \gate{R_x} & \targ & \gate{R_z} & \ctrl{1} & \gate{R_x} & \gate{H} & \qw & \qw && \cdots\\ 	 	\lstick{ {q}_{4} :  } & \gate{H} & \ctrl{1} & \gate{R_z} & \control\qw & \gate{R_x} & \ctrl{1} & \gate{R_z} & \control\qw & \gate{R_x} & \ctrl{1} & \gate{R_z} & \control\qw & \gate{R_x} & \ctrl{1} & \gate{R_z} & \control\qw & \gate{R_x} & \gate{H} & \qw & \qw && \cdots\\ 	 	\lstick{ {q}_{5} :  } & \gate{H} & \targ & \gate{R_z} & \gate{R_x} & \qw & \targ & \gate{R_z} & \gate{R_x} & \qw & \targ & \gate{R_z} & \gate{R_x} & \qw & \targ & \gate{R_z} & \gate{R_x} & \gate{H} & \qw & \qw & \qw && \cdots\\ 	 	\lstick{\vdots} & \vdots & \vdots & \vdots & \vdots & \vdots & \vdots & \vdots & \vdots & \vdots & \vdots & \vdots & \vdots & \vdots & \vdots & \vdots & \vdots & \vdots & \vdots && \vdots \\ \\	 }
	\end{cases}}_{k}
\end{equation*}
\centering{}\caption{Test quantum circuit for $n$ qubits and rounds $k$.\label{fig:Quantum-Circuit-representing-1-1}}
\end{figure}

We use \texttt{quimb} \cite{Gray2018} to specify quantum circuits
and generate TNs therefrom. In particular we simulate circuits for
various $n$ qubits and \textit{rounds} $k$ where each consists of
alternating qubit couplings according of the form in figure \ref{fig:Quantum-Circuit-representing-1-1}.
We also use Bayesian parameter optimization (BPO) \cite{Gray_2021}
to find tensor contraction orders and compare against naive greedy
search. Note that for both strategy we set a timeout of 600 seconds.
We then deploy the contraction strategy that produces the fewest number
of tensor contractions balanced against the orders of intermediate
tensors\footnote{This choice was purely due to platform constraints in that large intermediate
tensors could not be effectively simulated.}. In order to expedite the process of deploying we precompute some
first few tensor contraction such that all tensors deployed to the
FPGA are square and congruent (i.e. all of the same dimensions). For
tensors of order greater than $\left(1,1\right)$ (i.e. tensors that
are not matrices) we transform them into $\left(1,1\right)$ tensors
by taking the Kronecker product of all component $\left(1,1\right)$
tensors; to be precise we perform the following operation on the $\left(p,q\right)$
tensor 
\begin{algorithm}[H]
\inputencoding{latin9}\begin{lstlisting}[basicstyle={\ttfamily}]
    mats = [t[idx] for idx in np.ndindex(t.shape[:-2])]
    block_mat = block_diag(*qubit_mats)
\end{lstlisting}
\inputencoding{utf8}\end{algorithm}

\noindent where \texttt{block\_diag }builds a block diagonal matrix
of its arguments. All of our code has been made available on GitHub\footnote{\url{https://github.com/makslevental/fpga_stuff/} on the \texttt{complexmatmul}
branch.}.

\begin{figure*}
\subfloat[Output stationary systolic array processing element. \label{fig:os-matrix-multiplication-1}]{\begin{centering}
\includegraphics[width=1\linewidth]{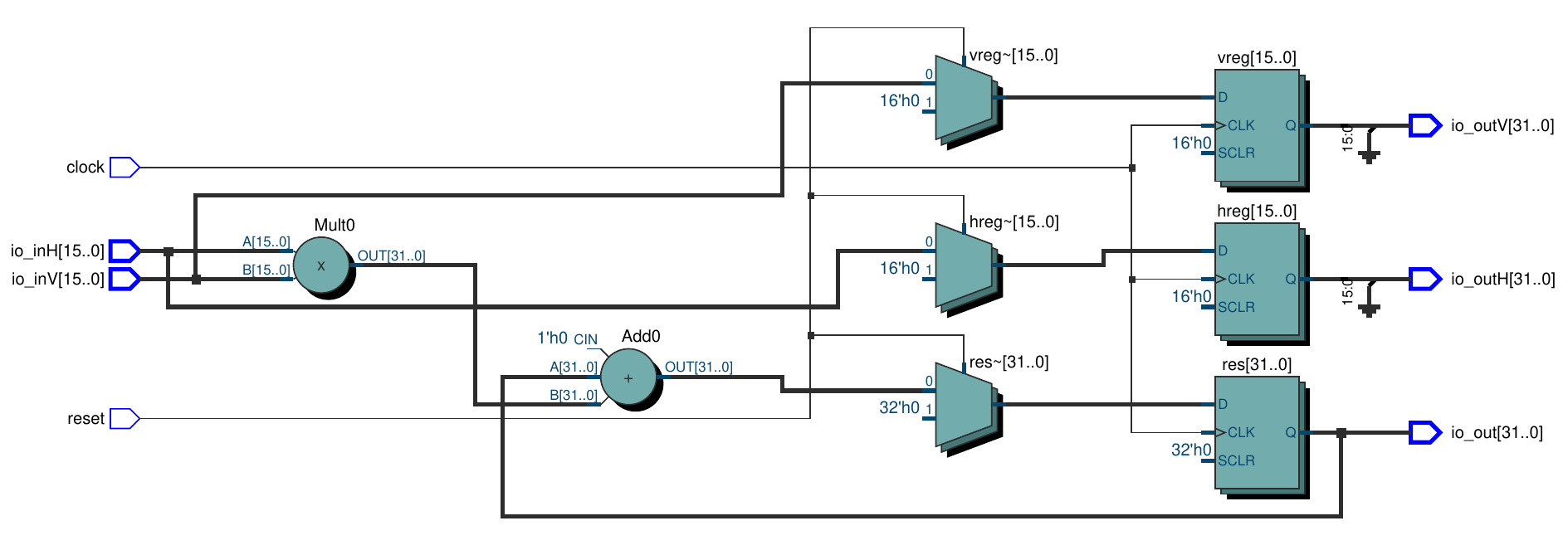}
\par\end{centering}
}\smallskip{}
\subfloat[$4\times4$ matrix multiplication systolic array implementation. \label{fig:os-matrix-multiplication}]{\begin{centering}
\includegraphics[width=1\linewidth]{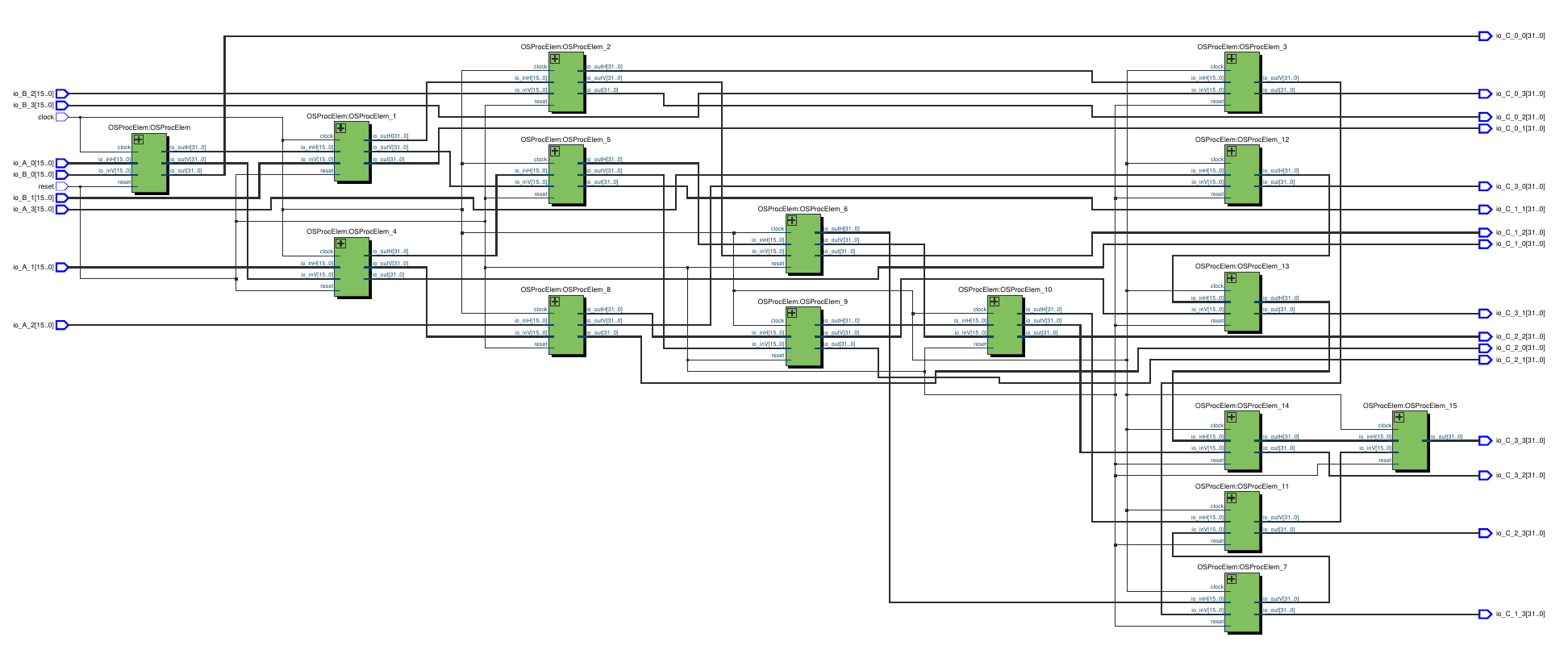}
\par\end{centering}
}

\smallskip{}

\noindent \raggedright{}\subfloat[Naive $4\times4$ matrix multiplication implementation. \label{fig:Naive--matrix}]{\begin{centering}
\includegraphics[angle=90,width=1\linewidth]{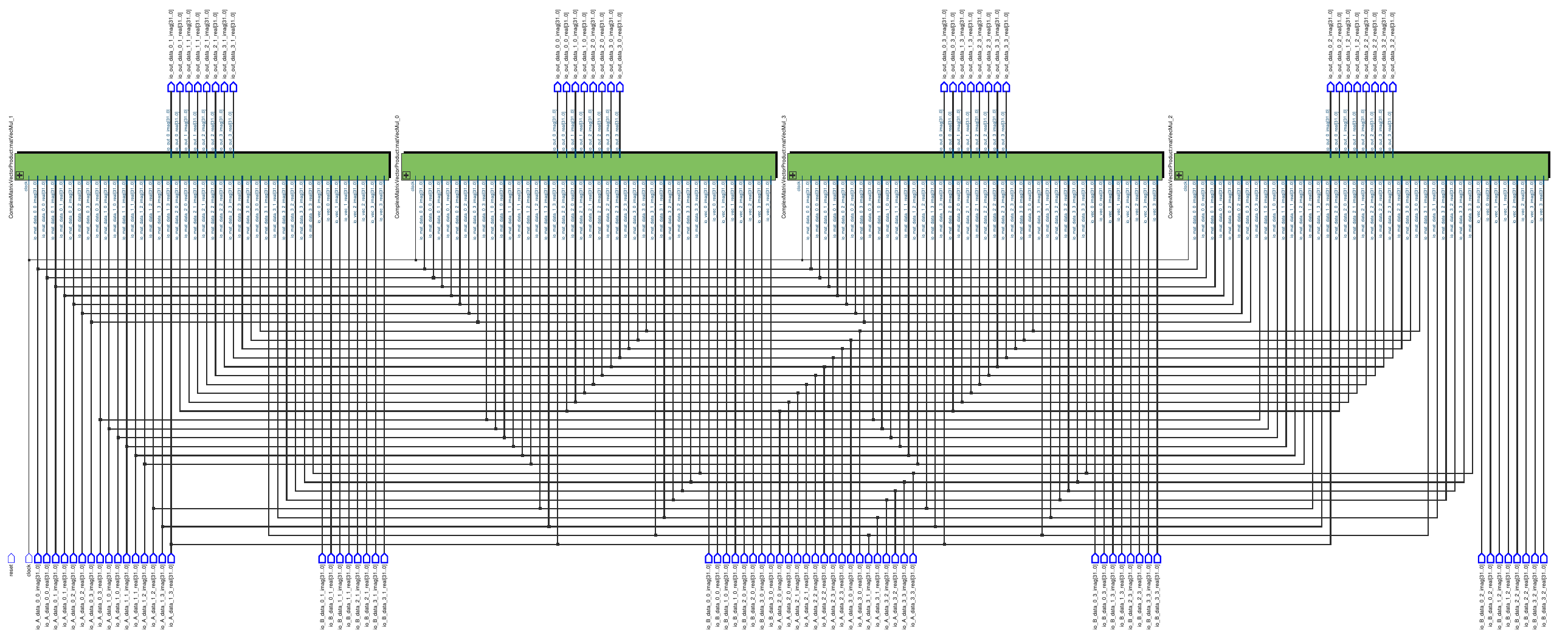}
\par\end{centering}
}\caption{FPGA implementations \label{fig:FPGA-implementations}}
\end{figure*}

For deploying circuits to FPGAs we use Chisel \cite{6241660} as a
HDL, by way of an adaptation of the Gemmini systolic array generator
\cite{genc2019gemmini}. Notably we experiment with using Gemmini
as an accelerator (i.e. fully parameterizing the weights/entries of
the matrices) and ``hardcoding'' certain gates/tensors. One possible
advantage of the latter approach over the former is a reduction in
loads from memory for the weights. The success of the chosen approach
depends heavily on whether certain sequences of fixed gates can actually
be pipelined or alternatively deployed in toto to the FPGA. We hypothesize
that this might depend on the depth and gate count of the circuits/TNs.
See figure \ref{fig:FPGA-implementations} for the netlists corresponding
to our systolic array and matmul implementations. Note that (complex)
arithmetic was done in 32 bit fixed precision for both implementations,
with 28 bits allocated behind the binary point. 

One challenge we faced was in deploying to real hardware\footnote{A challenge not unfamiliar to the seasoned QC researcher.};
unfortunately time and administrative challenges\footnote{We were not able to get allocations on CHI@TACC in a timely fashion
(the issue is ongoing...).} prevented us from actually deploying to real FPGAs. As a substitute
we used the well-known and trusted Verilog simulator\footnote{It is simulations all the way down.}
Verilator\footnote{\url{https://www.veripool.org/verilator/}}, which
transpiles Verilog (which Chisel generates) to a cycle-accurate model
in C++. We then executed this model to collect proxy measurements.
Note that for certain configurations (generally those with high qubit
and round count) we could successfully simulate due to memory constraints
on the workstation running the Verilator produced model.

\section{Evaluation\label{sec:Evaluation}}

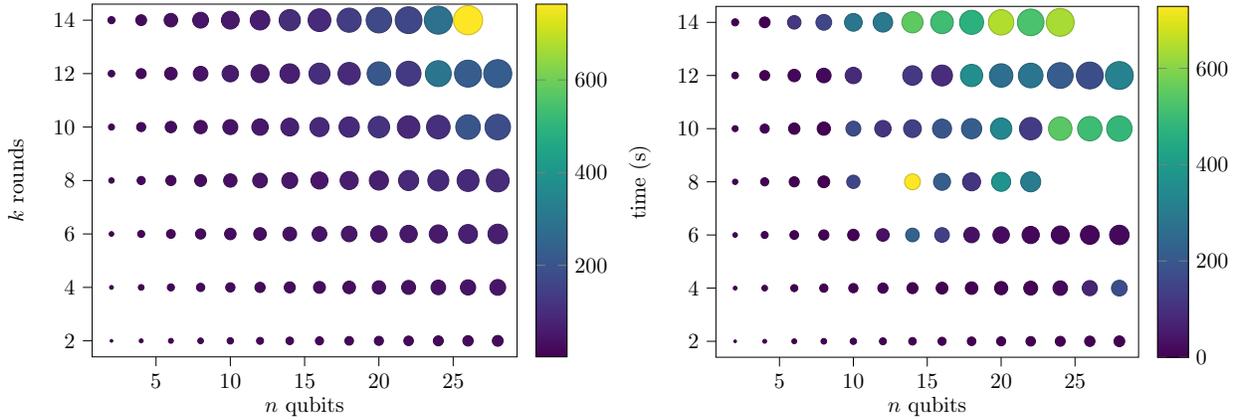
\begin{figure*}
\centering{}\subfloat[Runtimes for computing BPO contraction strategy.\label{fig:opt_contract_run-1}]{\centering{}\noindent\resizebox{.5\linewidth}{!}{% This file was created by tikzplotlib v0.9.8.
\begin{tikzpicture}

\begin{axis}[
colorbar,
colorbar style={ylabel={}},
colormap/viridis,
point meta max=763.48309636116,
point meta min=2.55694222450256,
tick align=outside,
tick pos=left,
x grid style={white!69.0196078431373!black},
xlabel={\(\displaystyle n\) qubits},
xmin=0.7, xmax=29.3,
xtick style={color=black},
y grid style={white!69.0196078431373!black},
ylabel={\(\displaystyle k\) rounds},
ymin=1.4, ymax=14.6,
ytick style={color=black}
]
\addplot [
  colormap/viridis,
  only marks,
  scatter,
  scatter src=explicit,
  scatter/@pre marker code/.append style={/tikz/mark size=\perpointmarksize},
  visualization depends on={\thisrow{sizedata} \as\perpointmarksize}
]
table [x=x, y=y, meta=colordata]{%
x  y  colordata  sizedata
2 2 2.5569422245025635 0.6180387232371033
4 2 5.792724847793579 0.909728368293446
6 2 8.053183794021606 1.1283791670955126
8 2 8.64632534980774 1.3110581167104949
10 2 10.161460399627686 1.4712264360219254
12 2 11.881834030151367 1.6155931006002358
14 2 11.76687240600586 1.7480774889473265
16 2 13.745242595672607 1.8712051592547776
18 2 13.922746896743774 1.9867165345562021
20 2 14.878172636032104 2.095871281671733
22 2 16.767428874969482 2.199615936929358
24 2 16.846333265304565 2.298683125324351
26 2 19.75782155990601 2.393653682408596
28 2 21.0377094745636 2.4849973424463734
2 4 3.98638653755188 0.8740387444736633
4 4 9.863959074020386 1.3110581167104949
6 4 15.06441068649292 1.6351767622932518
8 4 17.033417463302612 1.9049232799499338
10 4 20.37151837348938 2.140948939383325
12 4 22.31747317314148 2.3534213434057993
14 4 23.82140278816223 2.548238936628457
16 4 26.34659099578857 2.729185104880338
18 4 27.844101905822757 2.8988585676524603
20 4 31.33110213279724 3.0591356056578216
22 4 35.153897523880005 3.211423409074988
24 4 37.38166379928589 3.3568094927931478
26 4 39.790363788604736 3.496154977894653
28 4 45.592653036117554 3.6301555459799424
2 6 7.495872735977173 1.0704744696916626
4 6 12.41064715385437 1.6155931006002358
6 6 18.897825956344604 2.018506017616128
8 6 22.436465740203857 2.3534213434057993
10 6 25.62346887588501 2.6462837142006137
12 6 29.89509105682373 2.9098183744847086
14 6 34.798293352127075 3.1513915099419605
16 6 39.81081581115723 3.3757212452126
18 6 46.73789143562317 3.5860450919955182
20 6 45.498624086380005 3.7846987830302403
22 6 53.27591323852539 3.9734330691124042
24 6 54.56215238571167 4.153600345623235
26 6 67.67761278152466 4.326271062659724
28 6 66.17209959030151 4.492309738213999
2 8 7.713158130645752 1.2360774464742066
4 8 15.74844241142273 1.8712051592547776
6 8 21.40258836746216 2.339856842279288
8 8 26.48204112052917 2.729185104880338
10 8 31.34635329246521 3.0695231927842155
12 8 42.87822270393372 3.3757212452126
14 8 46.7872154712677 3.656366395715726
16 8 46.96520471572876 3.916955005365611
18 8 60.88023710250855 4.161256758288079
20 8 63.74074029922485 4.391990335000489
22 8 87.30655860900879 4.611192947283514
24 8 89.60698556900024 4.820437914901168
26 8 89.73227167129517 5.0209703231304035
28 8 96.16073298454285 5.21379557329358
2 10 10.105061531066896 1.381976597885342
4 10 18.448184967041016 2.095871281671733
6 10 24.748371124267575 2.6221162334209898
8 10 32.76072645187378 3.0591356056578216
10 10 38.44206142425537 3.441093978088511
12 10 53.11485123634338 3.7846987830302403
14 10 68.39667820930481 4.099605101775554
16 10 90.27946162223816 4.391990335000489
18 10 93.86561036109924 4.666090035026252
20 10 113.55598831176758 4.924958205630262
22 10 86.57283782958984 5.170882945826411
24 10 103.03463411331175 5.405631096520737
26 10 201.88830542564392 5.630600737390776
28 10 178.60313725471497 5.846920708897899
2 12 11.167222023010254 1.5138795132120961
4 12 21.52735948562622 2.298683125324351
6 12 27.526602745056152 2.876813695875796
8 12 40.05171537399292 3.3568094927931478
10 12 58.41967535018921 3.7762789755305186
12 12 53.58670949935913 4.153600345623235
14 12 67.87978267669678 4.499389820996742
16 12 101.45704889297484 4.820437914901168
18 12 94.4800066947937 5.121399674068052
20 12 210.1106550693512 5.405631096520737
22 12 131.3973731994629 5.67564626115825
24 12 300.92115902900696 5.93338633597436
26 12 224.10146594047544 6.180387232371033
28 12 224.59903526306152 6.417889004352016
2 14 11.77914571762085 1.6351767622932518
4 14 20.974145889282227 2.4849973424463734
6 14 32.39659643173218 3.110726690017501
8 14 44.87882828712464 3.6301555459799424
10 14 53.11286354064941 4.084046772017999
12 14 62.34402441978455 4.492309738213999
14 14 100.22729635238647 4.866441773213158
16 14 91.47043824195862 5.21379557329358
18 14 138.4531877040863 5.539410891793226
20 14 168.10566234588623 5.846920708897899
22 14 164.70457315444946 6.139046385568431
24 14 284.2506809234619 6.417889004352016
26 14 763.4830963611603 6.68511092056102
};
\end{axis}

\end{tikzpicture}}}\subfloat[Runtimes for computing greedy contraction strategy.\label{fig:greedy_contract_run-1}]{\centering{}\noindent\resizebox{.5\linewidth}{!}{% This file was created by tikzplotlib v0.9.8.
\begin{tikzpicture}

\begin{axis}[
colorbar,
colorbar style={ylabel={}},
colormap/viridis,
point meta max=728.62079834938,
point meta min=0.0024678707122802,
tick align=outside,
tick pos=left,
x grid style={white!69.0196078431373!black},
xlabel={\(\displaystyle n\) qubits},
xmin=0.7, xmax=29.3,
xtick style={color=black},
y grid style={white!69.0196078431373!black},
ylabel={time (s)},
ymin=1.4, ymax=14.6,
ytick style={color=black}
]
\addplot [
  colormap/viridis,
  only marks,
  scatter,
  scatter src=explicit,
  scatter/@pre marker code/.append style={/tikz/mark size=\perpointmarksize},
  visualization depends on={\thisrow{sizedata} \as\perpointmarksize}
]
table [x=x, y=y, meta=colordata]{%
x  y  colordata  sizedata
2 2 0.0024678707122802 0.6180387232371033
4 2 0.0033321380615234 0.909728368293446
6 2 0.0055465698242187 1.1283791670955126
8 2 0.0072860717773437 1.3110581167104949
10 2 0.0089952945709228 1.4712264360219254
12 2 0.011221170425415 1.6155931006002358
14 2 0.0131881237030029 1.7480774889473265
16 2 0.0160574913024902 1.8712051592547776
18 2 0.0175197124481201 1.9867165345562021
20 2 0.0197861194610595 2.095871281671733
22 2 0.0236070156097412 2.199615936929358
24 2 0.0247032642364501 2.298683125324351
26 2 0.0287244319915771 2.393653682408596
28 2 0.0295631885528564 2.4849973424463734
2 4 0.0034406185150146 0.8740387444736633
4 4 0.0084962844848632 1.3110581167104949
6 4 0.0152878761291503 1.6351767622932518
8 4 0.0391011238098144 1.9049232799499338
10 4 0.1055092811584472 2.140948939383325
12 4 2.633755922317505 2.3534213434057993
14 4 0.0715532302856445 2.548238936628457
16 4 0.1337370872497558 2.729185104880338
18 4 0.4595546722412109 2.8988585676524603
20 4 1.5809521675109863 3.0591356056578216
22 4 5.057819128036499 3.211423409074988
24 4 18.03762197494507 3.3568094927931478
26 4 71.12972640991211 3.496154977894653
28 4 172.97193002700806 3.6301555459799424
2 6 0.0067255496978759 1.0704744696916626
4 6 0.0136914253234863 1.6155931006002358
6 6 0.0255858898162841 2.018506017616128
8 6 0.938269853591919 2.3534213434057993
10 6 1.9074623584747317 2.6462837142006137
12 6 32.96614146232605 2.9098183744847086
14 6 225.65068006515503 3.1513915099419605
16 6 136.6981999874115 3.3757212452126
18 6 31.31634783744812 3.5860450919955182
20 6 19.56786561012268 3.7846987830302403
22 6 11.373047351837158 3.9734330691124042
24 6 11.244412899017334 4.153600345623235
26 6 11.845577001571655 4.326271062659724
28 6 15.41252064704895 4.492309738213999
2 8 0.0068058967590332 1.2360774464742066
4 8 0.0257248878479003 1.8712051592547776
6 8 0.0596578121185302 2.339856842279288
8 8 0.4884233474731445 2.729185104880338
10 8 150.7149384021759 3.0695231927842155
14 8 728.6207983493805 3.656366395715726
16 8 215.90567326545715 3.916955005365611
18 8 107.89589238166808 4.161256758288079
20 8 371.75268602371216 4.391990335000489
22 8 303.75851225852966 4.611192947283514
2 10 0.0092372894287109 1.381976597885342
4 10 0.0413615703582763 2.095871281671733
6 10 0.3464555740356445 2.6221162334209898
8 10 1.9698970317840576 3.0591356056578216
10 10 166.64892148971558 3.441093978088511
12 10 108.77782201766968 3.7846987830302403
14 10 136.78280425071716 4.099605101775554
16 10 193.41166853904724 4.391990335000489
18 10 216.04517364501956 4.666090035026252
20 10 334.3462288379669 4.924958205630262
22 10 128.2732949256897 5.170882945826411
24 10 542.9363141059875 5.405631096520737
26 10 500.4453089237213 5.630600737390776
28 10 483.63575196266174 5.846920708897899
2 12 0.0110440254211425 1.5138795132120961
4 12 0.0740301609039306 2.298683125324351
6 12 11.461396932601929 2.876813695875796
8 12 14.559224128723145 3.3568094927931478
10 12 87.62356305122375 3.7762789755305186
14 12 119.37795996665956 4.499389820996742
16 12 92.45745420455933 4.820437914901168
18 12 359.1897826194763 5.121399674068052
20 12 260.8611209392548 5.405631096520737
22 12 281.9398362636566 5.67564626115825
24 12 211.9070816040039 5.93338633597436
26 12 172.46440505981445 6.180387232371033
28 12 321.25200748443604 6.417889004352016
2 14 0.012855052947998 1.6351767622932518
4 14 0.1670784950256347 2.4849973424463734
6 14 114.383971452713 3.110726690017501
8 14 154.7626233100891 3.6301555459799424
10 14 286.3877143859863 4.084046772017999
12 14 292.7816233634949 4.492309738213999
14 14 549.2970037460327 4.866441773213158
16 14 503.4743330478668 5.21379557329358
18 14 465.6411736011505 5.539410891793226
20 14 644.8817083835602 5.846920708897899
22 14 521.176038980484 6.139046385568431
24 14 634.9967432022095 6.417889004352016
};
\end{axis}

\end{tikzpicture}}}\caption{Runtimes for computing\textbf{ }contraction strategies for circuits
for various $n$ qubits and rounds $k$. Color scale represents time
and marker size represents, qualitatively, the number of tensors in
the tensor network corresponding to the quantum circuit. Note that
absent markers correspond to searches that didn't converge. \label{fig:runtimes-1}}
\end{figure*}
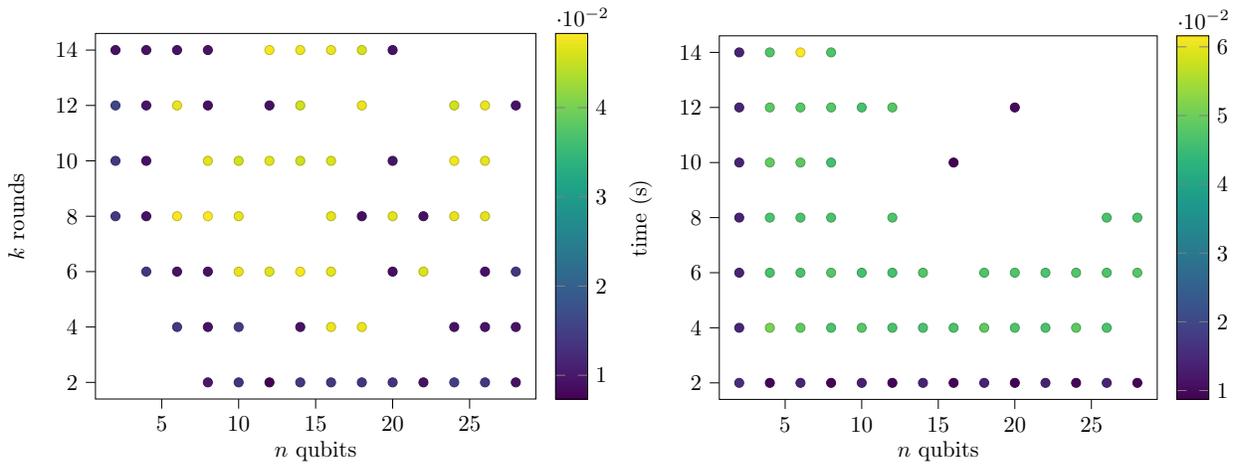
\begin{figure*}
\centering{}\subfloat[Runtimes for contracting with systolic array.\label{fig:opt_contract_run}]{\centering{}\noindent\resizebox{.5\linewidth}{!}{% This file was created by tikzplotlib v0.9.8.
\begin{tikzpicture}

\begin{axis}[
colorbar,
colorbar style={ylabel={}},
colormap/viridis,
point meta max=0.0483214,
point meta min=0.0073435,
tick align=outside,
tick pos=left,
x grid style={white!69.0196078431373!black},
xlabel={\(\displaystyle n\) qubits},
xmin=0.7, xmax=29.3,
xtick style={color=black},
y grid style={white!69.0196078431373!black},
ylabel={\(\displaystyle k\) rounds},
ymin=1.4, ymax=14.6,
ytick style={color=black}
]
\addplot [colormap/viridis, only marks, scatter, scatter src=explicit]
table [x=x, y=y, meta=colordata]{%
x  y  colordata
2 8 0.014041600000000001
2 10 0.013769600000000002
2 12 0.015611
2 14 0.009201999999999998
4 6 0.01370965
4 8 0.009170599999999998
4 10 0.009034999999999998
4 12 0.009337900000000001
4 14 0.009318799999999999
6 4 0.01380215
6 6 0.0093112
6 8 0.048068999999999994
6 12 0.0470719
6 14 0.009306699999999998
8 2 0.0098659
8 4 0.009144900000000001
8 6 0.0092672
8 8 0.0483214
8 10 0.046570799999999996
8 12 0.0089567
8 14 0.009395500000000001
10 2 0.013566599999999998
10 4 0.014333050000000002
10 6 0.0468923
10 8 0.046985
10 10 0.0455763
12 2 0.0073435
12 6 0.0464031
12 10 0.046626799999999996
12 12 0.0090037
12 14 0.0480242
14 2 0.013781550000000002
14 4 0.0093156
14 6 0.047868999999999995
14 10 0.0454605
14 12 0.045375700000000005
14 14 0.0474879
16 2 0.01375595
16 4 0.047262700000000005
16 6 0.04679610000000001
16 8 0.0467371
16 10 0.046142
16 14 0.047298200000000006
18 2 0.013555849999999998
18 4 0.0471931
18 8 0.0088638
18 12 0.04721299999999999
18 14 0.0458865
20 2 0.0132719
20 6 0.0088702
20 8 0.046558
20 10 0.0094773
20 14 0.0091107
22 2 0.0091275
22 6 0.046277
22 8 0.009069299999999999
24 2 0.015064866666666664
24 4 0.0090818
24 8 0.046919100000000005
24 10 0.0475875
24 12 0.04551809999999999
26 2 0.014012150000000001
26 4 0.0092823
26 6 0.0092419
26 8 0.046721599999999995
26 10 0.046856600000000005
26 12 0.047094899999999995
28 2 0.009108799999999997
28 4 0.009212499999999998
28 6 0.0136726
28 12 0.0092231
};
\end{axis}

\end{tikzpicture}}}\hspace*{\fill}\subfloat[Runtimes for contracting with matmul.\label{fig:greedy_contract_run}]{\centering{}\noindent\resizebox{.5\linewidth}{!}{% This file was created by tikzplotlib v0.9.8.
\begin{tikzpicture}

\begin{axis}[
colorbar,
colorbar style={ylabel={}},
colormap/viridis,
point meta max=0.0616001,
point meta min=0.0087104,
tick align=outside,
tick pos=left,
x grid style={white!69.0196078431373!black},
xlabel={\(\displaystyle n\) qubits},
xmin=0.7, xmax=29.3,
xtick style={color=black},
y grid style={white!69.0196078431373!black},
ylabel={time (s)},
ymin=1.4, ymax=14.6,
ytick style={color=black}
]
\addplot [colormap/viridis, only marks, scatter, scatter src=explicit]
table [x=x, y=y, meta=colordata]{%
x  y  colordata
2 2 0.015495566666666665
2 4 0.013989350000000001
2 6 0.013740950000000002
2 8 0.013808049999999999
2 10 0.0137095
2 12 0.0139794
2 14 0.013835
4 2 0.0090867
4 4 0.05080520000000001
4 6 0.0464044
4 8 0.0466712
4 10 0.048891500000000004
4 12 0.0481764
4 14 0.047275000000000005
6 2 0.0139827
6 4 0.0485813
6 6 0.046500400000000004
6 8 0.0465955
6 10 0.0480405
6 12 0.0478006
6 14 0.06160009999999999
8 2 0.009261799999999999
8 4 0.0466416
8 6 0.04784
8 8 0.046665
8 10 0.0459605
8 12 0.04776849999999999
8 14 0.0472837
10 2 0.0132136
10 4 0.047919100000000006
10 6 0.046026199999999996
10 12 0.0462656
12 2 0.009119799999999999
12 4 0.046374000000000005
12 6 0.046118599999999996
12 8 0.0470455
12 12 0.047656699999999996
14 2 0.01371665
14 4 0.0463552
14 6 0.0475169
16 2 0.009189300000000001
16 4 0.046451
16 10 0.008710400000000002
18 2 0.01385195
18 4 0.04854970000000001
18 6 0.047036699999999994
20 2 0.0087724
20 4 0.047033599999999995
20 6 0.0470217
20 12 0.0098028
22 2 0.013486149999999999
22 4 0.0464897
22 6 0.0466608
24 2 0.0093269
24 4 0.04852820000000001
24 6 0.04739120000000001
26 2 0.013852749999999999
26 4 0.046646499999999994
26 6 0.0469076
26 8 0.0464272
28 2 0.009089200000000002
28 6 0.047786
28 8 0.0466899
};
\end{axis}

\end{tikzpicture}}}\caption{Runtimes for test circuits for various $n$ qubits and rounds $k$.
Note that absent markers correspond to contractions that could not
be simulated due to memory constraints imposed by using the Verilator
model rather than an actual FPGA.\label{fig:runtimes}}
\end{figure*}
We perform two sets of evaluations. Even though it was not the central
goal of our exploration we first compare the time required to compute
a tensor contraction strategy across $n$ qubits and rounds $k$ for
the greedy search strategy and the BPO search strategy. We then address
our central goal in comparing the actual runtime for performing the
discovered tensor contraction on both the systolic array implementation
(see fig. \ref{fig:os-matrix-multiplication}) and the ``hardcoded''
naive matmul (see fig. \ref{fig:Naive--matrix}). 

Some interesting things to note searching for contractions: computing
(not evaluating) the optimal contraction strategy (i.e. using BPO)
is generally more performant that greedy search (see figures \ref{fig:opt_contract_run-1},
\ref{fig:greedy_contract_run-1}). The likely reason for this is that
BPO converges more quickly and more efficiently searches the space
of possible contraction orders than greedy search (which greedily
optimizes some surrogate objective). Also note that, in fact, for
certain configurations greedy didn't converge within the timeout. 

Regarding the differences in the evaluation times of the contraction
orders (figures \ref{fig:opt_contract_run}, \ref{fig:greedy_contract_run})
it's clear that the systolic array implementation is more performant
in terms of both memory requirements and runtime. This is paradoxically
both obvious and suprising. As already mentioned, one expects systolic
arrays to have improved performance relative to naive matrix multiplication
for streaming data (and indeed, in general, they do) but for this
use case (where all matrix elements are known at deploy time) one
also expects that latency incurred by pipelining would offset that
performance improvement. One hypothesis for this is that the difference
is an artifact of simulating the FPGA implementations insofar as simulating
a more densely connected FPGA implementation (see the differences
between \ref{fig:os-matrix-multiplication} and \ref{fig:Naive--matrix})
is more compute intensive, especially with respect to heap allocations
(since systolic arrays incur more loads from memory). To corroborate
this hypothesis we used Intel's Quartus EDA\footnote{Electronic design automation.}
tool to synthesize and place and route (for an Arria II GX). Indeed
the naive implementation an order of magnitude (sometimes several)
of each type of resource (see table \ref{tab:Synthesis-and-place}).
\begin{table}

\begin{centering}
\begin{tabular}{ccc}
 & Systolic & Naive\tabularnewline
\midrule
LUTs & 512 & 110,074\tabularnewline
Registers & 896 & 2,048\tabularnewline
Pins & 642 & 3,074\tabularnewline
DSPs & 32 & 232\tabularnewline
\end{tabular}
\par\end{centering}

\begin{centering}
\caption{Synthesis and place and route for an Arria II GX for both systolic
arrays and naive matmul. \label{tab:Synthesis-and-place}}
\par\end{centering}
\end{table}

\section{Conclusion\label{sec:Conclusion}}

We explored tensor networks deployed to FPGAs as a mean of accelerating
simulations of quantum circuits. In order to accomplish this goal
we expressed tensor contraction as sequences of matrix multiplication
and implemented two different matrix multiplication FPGA designs:
systolic arrays, which operate on streaming matrix elements and naive
matrix multiplication, which wholesale instantiates all the necessary
MAC operations. In order to choose the contraction orders we used
an ``off the shelf'' library which searches for a suitable contraction
by either performing greedy search or Bayesian optimization. We compared
the performance of both the contraction search strategy and each contraction
evaluation implementation. Unfortunately we were unable to obtain
acecss to FPGA devices and thus we made due with cycle-accurate simulations.
Results for both comparison were generally in agreement with intuition:
BPO converged to a contraction order more effectively (more quickly
and more robustly) than greedy search and systolic arrays evaluated
the contraction more efficiently than naive matrix multiplication.

Possible future work includes actually deploying to real FPGAs and
then further comparing performance to the simulations performed here.
Another particularly interesting research direction is the tangential
problem of discovering optimal tensor contraction orders. Finding
such tensor contraction orders is ultimately a combinatorial optimization
problem. It occurs to us that possibly a deep learning approach could
be effective. Recently there has been work on RL for combinatorial
optimization\cite{Barrett_Clements_Foerster_Lvovsky_2020} and MCTS
for combinatorial optimization\cite{abe2020solving} that could, possibly,
be adapted to this problem in a straightforward fashion. 

\bibliographystyle{plain}
\phantomsection\addcontentsline{toc}{section}{\refname}\nocite{*}
\bibliography{biblio}

\begin{thebibliography}{10}

\bibitem{8015156}
K.~Abdelouahab, M.~Pelcat, J.~Sérot, C.~Bourrasset, and F.~Berry.
\newblock Tactics to directly map cnn graphs on embedded fpgas.
\newblock {\em IEEE Embedded Systems Letters}, 9(4):113--116, 2017.

\bibitem{abe2020solving}
Kenshin Abe, Zijian Xu, Issei Sato, and Masashi Sugiyama.
\newblock Solving np-hard problems on graphs with extended alphago zero, 2020.

\bibitem{Arnborg87complexityof}
Stefan Arnborg, Derek~G. Corneil, and Andrzej Proskurowski.
\newblock Complexity of finding embeddings in a k-tree.
\newblock {\em SIAM JOURNAL OF DISCRETE MATHEMATICS}, 8(2):277--284, 1987.

\bibitem{6241660}
Jonathan Bachrach, Huy Vo, Brian Richards, Yunsup Lee, Andrew Waterman, Rimas
  Avižienis, John Wawrzynek, and Krste Asanović.
\newblock Chisel: Constructing hardware in a scala embedded language.
\newblock In {\em DAC Design Automation Conference 2012}, pages 1212--1221,
  2012.

\bibitem{Barrett_Clements_Foerster_Lvovsky_2020}
Thomas Barrett, William Clements, Jakob Foerster, and Alex Lvovsky.
\newblock Exploratory combinatorial optimization with reinforcement learning.
\newblock {\em Proceedings of the AAAI Conference on Artificial Intelligence},
  34(04):3243--3250, Apr. 2020.

\bibitem{betelu2020limits}
Santiago~I. Betelu.
\newblock The limits of quantum circuit simulation with low precision
  arithmetic, 2020.

\bibitem{10.1145/3373087.3375296}
Johannes de~Fine~Licht, Grzegorz Kwasniewski, and Torsten Hoefler.
\newblock Flexible communication avoiding matrix multiplication on fpga with
  high-level synthesis.
\newblock In {\em Proceedings of the 2020 ACM/SIGDA International Symposium on
  Field-Programmable Gate Arrays}, FPGA '20, page 244–254, New York, NY, USA,
  2020. Association for Computing Machinery.

\bibitem{evenbly2019tensortrace}
Glen Evenbly.
\newblock Tensortrace: an application to contract tensor networks, 2019.

\bibitem{farhi2014quantum}
Edward Farhi, Jeffrey Goldstone, and Sam Gutmann.
\newblock A quantum approximate optimization algorithm, 2014.

\bibitem{feynman1982simulating}
Richard~P Feynman.
\newblock Simulating physics with computers.
\newblock {\em International journal of theoretical physics}, 21(6/7):467--488,
  1982.

\bibitem{Fried_2018}
E.~Schuyler Fried, Nicolas P.~D. Sawaya, Yudong Cao, Ian~D. Kivlichan,
  Jhonathan Romero, and Alán Aspuru-Guzik.
\newblock qtorch: The quantum tensor contraction handler.
\newblock {\em PLOS ONE}, 13(12):e0208510, Dec 2018.

\bibitem{genc2019gemmini}
Hasan Genc, Ameer Haj-Ali, Vighnesh Iyer, Alon Amid, Howard Mao, John Wright,
  Colin Schmidt, Jerry Zhao, Albert Ou, Max Banister, Yakun~Sophia Shao,
  Borivoje Nikolic, Ion Stoica, and Krste Asanovic.
\newblock Gemmini: An agile systolic array generator enabling systematic
  evaluations of deep-learning architectures, 2019.

\bibitem{glasser2019expressive}
Ivan Glasser, Ryan Sweke, Nicola Pancotti, Jens Eisert, and J.~Ignacio Cirac.
\newblock Expressive power of tensor-network factorizations for probabilistic
  modeling, with applications from hidden markov models to quantum machine
  learning, 2019.

\bibitem{Gray2018}
Johnnie Gray.
\newblock quimb: A python package for quantum information and many-body
  calculations.
\newblock {\em Journal of Open Source Software}, 3(29):819, 2018.

\bibitem{Gray_2021}
Johnnie Gray and Stefanos Kourtis.
\newblock Hyper-optimized tensor network contraction.
\newblock {\em Quantum}, 5:410, Mar 2021.

\bibitem{10.1145/237814.237866}
Lov~K. Grover.
\newblock A fast quantum mechanical algorithm for database search.
\newblock In {\em Proceedings of the Twenty-Eighth Annual ACM Symposium on
  Theory of Computing}, STOC '96, page 212–219, New York, NY, USA, 1996.
  Association for Computing Machinery.

\bibitem{j2020quantum}
Abhijith J., Adetokunbo Adedoyin, John Ambrosiano, Petr Anisimov, Andreas
  Bärtschi, William Casper, Gopinath Chennupati, Carleton Coffrin, Hristo
  Djidjev, David Gunter, Satish Karra, Nathan Lemons, Shizeng Lin, Alexander
  Malyzhenkov, David Mascarenas, Susan Mniszewski, Balu Nadiga, Daniel
  O'Malley, Diane Oyen, Scott Pakin, Lakshman Prasad, Randy Roberts, Phillip
  Romero, Nandakishore Santhi, Nikolai Sinitsyn, Pieter~J. Swart, James~G.
  Wendelberger, Boram Yoon, Richard Zamora, Wei Zhu, Stephan Eidenbenz,
  Patrick~J. Coles, Marc Vuffray, and Andrey~Y. Lokhov.
\newblock Quantum algorithm implementations for beginners, 2020.

\bibitem{9099977}
Liancheng Jia, Liqiang Lu, Xuechao Wei, and Yun Liang.
\newblock Generating systolic array accelerators with reusable blocks.
\newblock {\em IEEE Micro}, 40(4):85--92, 2020.

\bibitem{Kl_mper_1993}
A~Klümper, A~Schadschneider, and J~Zittartz.
\newblock Matrix product ground states for one-dimensional spin-1 quantum
  antiferromagnets.
\newblock {\em Europhysics Letters (EPL)}, 24(4):293–297, Nov 1993.

\bibitem{1653825}
Kung.
\newblock Why systolic architectures?
\newblock {\em Computer}, 15(1):37--46, 1982.

\bibitem{10.1137/050644756}
Igor~L. Markov and Yaoyun Shi.
\newblock Simulating quantum computation by contracting tensor networks.
\newblock {\em SIAM J. Comput.}, 38(3):963–981, June 2008.

\bibitem{McCaskey_2018}
Alexander McCaskey, Eugene Dumitrescu, Mengsu Chen, Dmitry Lyakh, and Travis
  Humble.
\newblock Validating quantum-classical programming models with tensor network
  simulations.
\newblock {\em PLOS ONE}, 13(12):e0206704, Dec 2018.

\bibitem{8764458}
Thierry Moreau, Tianqi Chen, Luis Vega, Jared Roesch, Eddie Yan, Lianmin Zheng,
  Josh Fromm, Ziheng Jiang, Luis Ceze, Carlos Guestrin, and Arvind
  Krishnamurthy.
\newblock A hardware–software blueprint for flexible deep learning
  specialization.
\newblock {\em IEEE Micro}, 39(5):8--16, 2019.

\bibitem{9103284}
K.~E. Murray, M.~A. Elgammal, V.~Betz, T.~Ansell, K.~Rothman, and A.~Comodi.
\newblock Symbiflow and vpr: An open-source design flow for commercial and
  novel fpgas.
\newblock {\em IEEE Micro}, 40(04):49--57, jul 2020.

\bibitem{10.1145/3020078.3021740}
Eriko Nurvitadhi, Ganesh Venkatesh, Jaewoong Sim, Debbie Marr, Randy Huang,
  Jason Ong Gee~Hock, Yeong~Tat Liew, Krishnan Srivatsan, Duncan Moss, Suchit
  Subhaschandra, and Guy Boudoukh.
\newblock Can fpgas beat gpus in accelerating next-generation deep neural
  networks?
\newblock In {\em Proceedings of the 2017 ACM/SIGDA International Symposium on
  Field-Programmable Gate Arrays}, FPGA '17, page 5–14, New York, NY, USA,
  2017. Association for Computing Machinery.

\bibitem{NAP25196}
National~Academies of~Sciences~Engineering and Medicine.
\newblock {\em Quantum Computing: Progress and Prospects}.
\newblock The National Academies Press, Washington, DC, 2019.

\bibitem{pednault2020paretoefficient}
Edwin Pednault, John~A. Gunnels, Giacomo Nannicini, Lior Horesh, Thomas
  Magerlein, Edgar Solomonik, Erik~W. Draeger, Eric~T. Holland, and Robert
  Wisnieff.
\newblock Pareto-efficient quantum circuit simulation using tensor contraction
  deferral, 2020.

\bibitem{roman2007advanced}
S.~Roman.
\newblock {\em Advanced Linear Algebra}.
\newblock Graduate Texts in Mathematics. Springer New York, 2007.

\bibitem{10.1007/978-3-319-94340-4_4}
Justin Sanchez, Nasim Soltani, Pratik Kulkarni, Ramachandra~Vikas Chamarthi,
  and Hamed Tabkhi.
\newblock A reconfigurable streaming processor for real-time low-power
  execution of convolutional neural networks at the edge.
\newblock In Shijun Liu, Bedir Tekinerdogan, Mikio Aoyama, and Liang-Jie Zhang,
  editors, {\em Edge Computing -- EDGE 2018}, pages 49--64, Cham, 2018.
  Springer International Publishing.

\bibitem{PhysRevA.74.022320}
Y.-Y. Shi, L.-M. Duan, and G.~Vidal.
\newblock Classical simulation of quantum many-body systems with a tree tensor
  network.
\newblock {\em Phys. Rev. A}, 74:022320, Aug 2006.

\bibitem{365700}
P.W. Shor.
\newblock Algorithms for quantum computation: discrete logarithms and
  factoring.
\newblock In {\em Proceedings 35th Annual Symposium on Foundations of Computer
  Science}, pages 124--134, 1994.

\bibitem{Verstraete:2004cf}
F.~Verstraete and J.~I. Cirac.
\newblock {Renormalization algorithms for quantum-many body systems in two and
  higher dimensions}.
\newblock 7 2004.

\bibitem{PhysRevLett.99.220405}
G.~Vidal.
\newblock Entanglement renormalization.
\newblock {\em Phys. Rev. Lett.}, 99:220405, Nov 2007.

\bibitem{10.1145/3431920.3439292}
Jie Wang, Licheng Guo, and Jason Cong.
\newblock Autosa: A polyhedral compiler for high-performance systolic arrays on
  fpga.
\newblock In {\em The 2021 ACM/SIGDA International Symposium on
  Field-Programmable Gate Arrays}, FPGA '21, page 93–104, New York, NY, USA,
  2021. Association for Computing Machinery.

\bibitem{warden_2015}
Pete Warden.
\newblock Why gemm is at the heart of deep learning, Apr 2015.

\bibitem{PhysRevLett.69.2863}
Steven~R. White.
\newblock Density matrix formulation for quantum renormalization groups.
\newblock {\em Phys. Rev. Lett.}, 69:2863--2866, Nov 1992.

\bibitem{10.1145/3295500.3356155}
Xin-Chuan Wu, Sheng Di, Emma~Maitreyee Dasgupta, Franck Cappello, Hal Finkel,
  Yuri Alexeev, and Frederic~T. Chong.
\newblock Full-state quantum circuit simulation by using data compression.
\newblock In {\em Proceedings of the International Conference for High
  Performance Computing, Networking, Storage and Analysis}, SC '19, New York,
  NY, USA, 2019. Association for Computing Machinery.

\bibitem{10.1007/978-3-030-05677-3_16}
Zhijie Yang, Lei Wang, Dong Ding, Xiangyu Zhang, Yu~Deng, Shiming Li, and Qiang
  Dou.
\newblock Systolic array based accelerator and algorithm mapping for deep
  learning algorithms.
\newblock In Feng Zhang, Jidong Zhai, Marc Snir, Hai Jin, Hironori Kasahara,
  and Mateo Valero, editors, {\em Network and Parallel Computing}, pages
  153--158, Cham, 2018. Springer International Publishing.

\bibitem{Zhong1460}
Han-Sen Zhong, Hui Wang, Yu-Hao Deng, Ming-Cheng Chen, Li-Chao Peng, Yi-Han
  Luo, Jian Qin, Dian Wu, Xing Ding, Yi~Hu, Peng Hu, Xiao-Yan Yang, Wei-Jun
  Zhang, Hao Li, Yuxuan Li, Xiao Jiang, Lin Gan, Guangwen Yang, Lixing You,
  Zhen Wang, Li~Li, Nai-Le Liu, Chao-Yang Lu, and Jian-Wei Pan.
\newblock Quantum computational advantage using photons.
\newblock {\em Science}, 370(6523):1460--1463, 2020.

\end{thebibliography}

\end{document}